\documentclass[preprint,11pt,nofootinbib]{revtex4-1}
\usepackage{amsmath}
\usepackage{amsfonts}
\usepackage{amssymb}
\usepackage{mathtools}
\usepackage{etoolbox}
\usepackage{tikz}
\usetikzlibrary{shapes.geometric}
\usepackage{slashed}
\usepackage{subcaption}
\usepackage{braket}
\usepackage{listings}
\usepackage[dvipdfmx]{colortbl}
\usepackage{xcolor}
\usepackage{appendix}
\usepackage{hyperref}
\usepackage{enumitem}
\usepackage{diagbox} 
\usepackage{listings}
\usepackage{float}
\usepackage{multirow}
\usepackage{lipsum}
\usepackage{array}
\usepackage{float}
\usepackage{bm}
\usepackage[left=35mm, top=30mm,bottom=33mm,right=25mm]{geometry}
\usetikzlibrary{calc,matrix,fit}
\DeclareMathOperator{\Tr}{Tr}

\newcommand{\be}{\begin{equation}}
\newcommand{\ee}{\end{equation}}
\newcommand{\bea}{\begin{eqnarray}}
\newcommand{\eea}{\end{eqnarray}}
\newcommand{\non}{\nonumber}
\newcommand{\bi}{\begin{itemize}}
\newcommand{\ei}{\end{itemize}}

 \DeclareMathOperator{\arccosh}{arccosh}

\begin{document}

\begin{flushright}
KANAZAWA-24-03  \\ UTHEP-786 \\ UTCCS-P-153
\end{flushright}

\title{
Spectroscopy with the tensor renormalization group method
}

\author{Fathiyya Izzatun Az-zahra$^1$}
\email{Contact author: fathiyya@hep.s.kanazawa-u.ac.jp}
\author{Shinji Takeda$^{1}$}
\email{Contact author: takeda@hep.s.kanazawa-u.ac.jp}
\author{Takeshi Yamazaki$^{2,3}$}
\email{Contact author: yamazaki@het.ph.tsukuba.ac.jp}

\affiliation{${}^1$\footnotesize{Institute for Theoretical Physics, Kanazawa University, Kanazawa 920-1192, Japan}\\
${}^{2}$\footnotesize{Institute of Pure and Applied Sciences, University of Tsukuba, Tsukuba 305-8571, Japan}\\
${}^{3}$\footnotesize {Center for Computational Sciences, University of Tsukuba, Tsukuba 305-8577, Japan}
}

\begin{abstract}
We present a spectroscopy scheme for the lattice field theory by using the tensor renormalization group method combining with the transfer matrix formalism.
By using the scheme, we cannot only compute the energy spectrum for the lattice theory but also determine quantum numbers of the energy eigenstates.
Furthermore, the wave function of the corresponding eigenstate can also be computed.
The first step of the scheme is to coarse grain the tensor network of a given lattice model by using the higher order tensor renormalization group, and then after making a matrix corresponding to a transfer matrix from the coarse-grained tensors, its eigenvalues are evaluated to extract the energy spectrum.
Second, the quantum number of the eigenstates can be identified by a selection rule that requires to compute matrix elements of an associated insertion operator.
The matrix elements can be represented by an impurity tensor network and computed by the coarse-graining scheme.
Moreover, we can compute the
wave function of the energy eigenstate by putting the impurity tensor at each point in space direction of the network.
Additionally, the momentum of the eigenstate can also be identified by computing appropriate matrix elements represented by the tensor network.
As a demonstration of the new scheme, we show the spectroscopy of the $(1+1)$d Ising model and compare it with exact results.
We also present a scattering phase shift obtained from two-particle state energy using L\"uscher's formula.
\end{abstract}

\maketitle

\tableofcontents

\section{INTRODUCTION}

Computing the energy spectrum and eigenstates is a fundamental and important task when studying a given quantum system.
For example, in lattice quantum chromodynamics (QCD),
where the Monte Carlo method is usually used, the hadron spectrum is obtained by computing the two-point
function of a given insertion operator that belongs to a desired quantum channel.
The methodology of the hadron spectroscopy has been well developed so far \cite{Wagner:2013tiz,Wittig2020}
 and
the numerical results are in good agreement with the experimental values \cite{Workman:2022ynf},
but there are unavoidable practical difficulties in the method.
For instance, when one wants to accurately obtain the lowest energy gap,
the Euclidean time extent should be taken quite large to suppress the effect of higher excited states.
Furthermore, if one wants to extract the energy spectrum of higher excited states,
very large statistics are required to suppress the statistical noise.
Motivated by these difficulties,
we look for alternative numerical tools for the spectroscopy.
A potential candidate is the tensor network method (see \cite{Banuls:2019rao,Okunishi:2021but,Kadoh:2022loj} for review) that
can be classified into two groups: Hamiltonian formalism \cite{PhysRevLett.69.2863,PhysRevLett.75.3537,Verstraete:2004cf,Banuls:2016gid,Banuls:2019bmf,Banuls:2019hzc,Schneider:2022lcl} and Lagrangian formalism \cite{
Shimizu:2012wfa,
Shimizu:2012zza,
Yu:2013sbi,
Zou:2014rha,
Shimizu:2014uva,
Shimizu:2014fsa,
Takeda:2014vwa,
Yang:2015rra,
Kawauchi:2016xng,
Shimizu:2017onf,
Kadoh:2018hqq,
Kadoh:2018tis,
Kuramashi:2018mmi,
Kuramashi:2019cgs,
Bazavov:2019qih,
Akiyama:2019xzy,
Akiyama:2020ntf,
Akiyama:2020soe,
Akiyama:2021zhf,
Akiyama:2021xxr,
Akiyama:2021glo,
Nakayama:2021iyp,
Akiyama:2022eip,
Akiyama:2023hvt,
Kuwahara:2022ubg,
Hirasawa:2021qvh,
Fukuma:2021cni,
Bloch:2021mjw,
Luo:2022eje,
Bloch:2022vqz,
Jha:2022pgy}.
For example, the spectroscopy using the former was done in
\cite{Banuls:2013jaa,Itou:2023img} for
$(1+1)$d quantum electrodynamics (QED).
On the other hand, the spectroscopy for the latter is discussed in \cite{PhysRevB.107.205123,PhysRevB.108.024413},
but the quantum number identification was not addressed.
In the current work we will complete the spectroscopy using the Lagrangian formalism and this is the main purpose of the paper.

Our new spectroscopy scheme starts by considering the transfer matrix formalism. 
In principle, a direct diagonalization of the transfer matrix provides us the exact energy spectrum of a system
and it does not require the large time extent in contrast to the Monte Carlo method.
The transfer matrix itself, however, has a very large dimensionality 
and it increases exponentially with respect to the volume of a system.
In order to reduce the dimensionality, we employ the tensor renormalization group (TRG) method
that uses the information compression technique based on the singular value decomposition.
So far many TRG coarse-graining algorithms are proposed \cite{PhysRevLett.99.120601,PhysRevB.86.045139,PhysRevLett.103.160601,
evenbly2015tensor,
yang2017loop,
Hauru:2017jbf,
morita2018tensor,
harada2018entanglement,
PhysRevB.99.155101,
PhysRevB.102.054432,
Kadoh:2019kqk,
Kadoh:2021fri,
Arai:2022uee,
Nakayama:2023ytr,homma2023},
but we here choose the higher order tensor renormalization group (HOTRG)  \cite{PhysRevB.86.045139} since
it has relatively high accuracy and can be extended into higher dimensional systems.
By using the new scheme, 
we are not only able to compute the energy spectrum
but also classify the quantum number of the energy eigenstates.
The latter procedure can be done by a selection rule that is derived from a symmetry of the system.
A crucial quantity in the selection rule is the matrix element of an interpolating operator associated with the symmetry.
The matrix element can be represented by the tensor network with some impurity and evaluated by the coarse-graining scheme.
Moreover, we can compute the wave function of the energy eigenstate
from the matrix element where a proper operator is inserted at each point in the space direction of the lattice.
From the position dependence of the wave function, we can infer the momentum of the state.
We will demonstrate the new scheme by applying to the $(1+1)$d Ising model and
show that the energy spectrum and the quantum number are correctly reproduced by
comparing with the exact results \cite{PhysRev.76.1232}.
Furthermore, we will show a scattering phase shift obtained from the two-particle state energy using L\"uscher's formula \cite{Luscher:1990ck,Gattringer:1992np,
Luscher:1985dn,Luscher:1986pf,LUSCHER1991531,Gockeler:1994rx,RUMMUKAINEN1995397,CP-PACS:2004dtj,
CP-PACS:2005gzm}.

The rest of the paper is organized as follows. 
Theoretical basics are summarized in Sec.~\ref{sec:formulation}. 
We briefly remind of the spectroscopy using the correlation function in Sec.~\ref{sec:spectroscopy} 
and the transfer matrix formalism in Sec.~\ref{sec:TM}.
In Sec.~\ref{sec:how_to}, 
we explain how to numerically obtain the energy spectrum 
and how to identify the quantum number of energy eigenstate by using the tensor renormalization group method 
and this is a key section of the paper. 
The numerical results for the $(1+1)$d Ising model are given in Secs.~\ref{sec:res} 
where the energy spectrum, the quantum number classification, momentum identification, and the scattering phase shift are
presented in Sec.~\ref{sec:energy_spectrum}, \ref{sec:quantum_number}, \ref{sec:momentum}, and \ref{sec:phase_shift}, respectively. 
The summary is given in the final section.
In Appendix \ref{sec:TN_TM_2DIsing}, we summarize the transfer matrix and the tensor network representation for the Ising model.
The exact spectrum of the transfer matrix for the Ising model is summarized in Appendix \ref{sec:exact_spectrum}.   
For comparison, a spectroscopy using the one-time slice transfer matrix is given in Appendix \ref{sec:1d_TN}.

\section{FORMULATION}
\label{sec:formulation} 

\subsection{Spectroscopy using correlation function}
\label{sec:spectroscopy}
Let us briefly remind how to obtain the energy spectrum from a correlation function \cite{Wagner:2013tiz,Wittig2020}.
In the continuum Euclidean space-time with time extent $T$, the correlation function for
an interpolating operator is defined as
\begin{equation}\label{correlator_def}
\langle \hat {\cal O}^\dag_q (\tau)\hat {\cal O}_q(0)\rangle
=
\frac{{\rm Tr}\left[\hat {\cal O}^\dag_q(\tau)\hat{\cal O}_q(0)e^{-\hat HT}\right]}{{\rm Tr} [e^{-\hat HT}]}
\end{equation}
where
$\hat {\cal O}_q(\tau)$ is the Euclidean time Heisenberg operator whose quantum number is denoted by $q$,
\begin{equation}
\hat {\cal O}_q(\tau)=e^{\hat H\tau}\hat {\cal O}_q(0)e^{-\hat H\tau}
\end{equation}
and $\hat H$ is the Hamiltonian of a system.
In a finite spatial volume, the eigenvalue of $\hat H$ is discretized
\begin{equation}
\hat H|n,q^\prime\rangle=E_{n,q^\prime}|n,q^\prime\rangle
\label{eqn:energy_eigen_equation}
\end{equation}
for $n=0,1,2,\hdots$ and for all possible quantum numbers $q^\prime$. 
The spectral decomposition of the numerator in the correlation function is given by
\begin{equation}
{\rm Tr}\left[\hat{\cal O}^\dag_q(\tau)\hat{\cal O}_q(0)e^{-\hat HT}\right]
=
\sum_{m,n=0}^\infty\sum_{q^\prime,q^{\prime\prime}}
|\langle n,q^{\prime\prime}|\hat {\cal O}_q(0)|m,q^\prime\rangle|^2
e^{-(T-\tau) E_{m,q^\prime}}e^{-\tau E_{n,q^{\prime\prime}}}.
\end{equation}

For the large $T$ limit, 
in the summation of $m$ and $q^\prime$,
the ground state, that is, the minimum energy eigenstate of the vacuum channel $(m,q^\prime)=(0,{\rm vac})$
dominates the summations
\be
{\rm Tr}\left[\hat{\cal O}^\dag_q(\tau)\hat{\cal O}_q(0)e^{-\hat HT}\right]
\stackrel{T\to\infty}{\sim}
\sum_{n=0}^\infty\sum_{q^{\prime\prime}}
|\langle n,q^{\prime\prime}|\hat{\cal O}_q(0)|\Omega\rangle|^2
e^{-(T-\tau) E_{\Omega}}e^{-\tau E_{n,q^{\prime\prime}}}
\ee
where the ground state is denoted by $|\Omega\rangle=|0,{\rm vac}\rangle$.
Furthermore, thanks to the conservation of quantum numbers, the
$q^{\prime\prime}=q$ sector only survives
\be
\langle n,q^{\prime\prime}|\hat{\cal O}_q(0)|\Omega\rangle
\neq
0
\hspace{10mm}
\mbox{ for }
q^{\prime\prime}=q
\ee
and other matrix elements vanish, thus we have
\be
{\rm Tr}\left[\hat{\cal O}^\dag_q(\tau)\hat{\cal O}_q(0)e^{-\hat HT}\right]
\stackrel{T\to\infty}{\sim}
\sum_{n=0}^\infty
|\langle n,q|\hat{\cal O}_q(0)|\Omega\rangle|^2
e^{-(T-\tau) E_{\Omega}}e^{-\tau E_{n,q}}.
\ee
In total, by taking into account the denominator,
the correlation function in the $T\to\infty$ limit is given by
\be
\lim_{T\to\infty}
\frac{
{\rm Tr}\left[\hat{\cal O}^\dag_q(\tau)\hat{\cal O}_q(0)e^{-\hat HT}\right]
}{
{\rm Tr}[e^{-\hat HT}]
}
=
\sum_{n=0}^\infty
|\langle n,q|\hat{\cal O}_q(0)|\Omega\rangle|^2
e^{-\tau(E_{n,q}-E_\Omega)}.
\ee
In a usual Monte Carlo (MC) simulation for the hadron spectroscopy, one computes the correlation function
with a proper interpolating operator for several time separations
then the energy gaps,
\be
\omega_{n,q}=E_{n,q}-E_\Omega
\label{eqn:energy_gap}
\ee
are extracted from the data.
Such a computation requires the large time extent as well as the large time separation to
avoid the contamination due to the higher excited states.
Furthermore, it is usually difficult to extract the energy of the higher excited states, thus one needs
sophisticated methods, say, the variational method \cite{Luscher:1990ck} and so on.

\subsection{Transfer matrix formalism for lattice field theory}
\label{sec:TM}
Needless to say, the computation of the correlation function is not the only way to obtain the energy spectrum.
A more direct method is the diagonalization of the Hamiltonian or equivalently the transfer matrix.
In fact, this method does not require computing the correlation function
or increasing the time extent in contrast to the MC calculation.
In this subsection, we briefly remind about the transfer matrix formalism
for lattice field theory.
In the following, the lattice spacing is set to $a=1$.

For simplicity, here we consider the field theory on the two-dimensional lattice,
although the discussion here can be straightforwardly extended to higher dimensional systems.
As a concrete example, we consider the lattice scalar field theory with the nearest-neighbor interaction
and a similar argument can be straightforwardly applied to fermion or gauge systems.
The scalar fields $\phi({\bm r})$ reside on the square lattice ${\bm r}=(t,x)\in\Gamma$ where the lattice $\Gamma$ is defined
\be
\Gamma=\{(t,x)|
t=0,1,2,\hdots,L_\tau-1
\,\,{\rm and}\,\,
x=0,1,2,\hdots, L_x-1
\}
\label{eqn:Gamma}
\ee
and the periodic boundary condition (PBC) is imposed on all directions.
Here the $0$ direction ($1$ direction) is considered as the time (space) direction.
The partition function of the system is given by
\be
Z=\int \prod_{{\bm r}\in\Gamma} d\phi({\bm r}) e^{-S[\phi]}
\ee
where the lattice action is given by
\be
S[\phi]=
\sum_{{\bm r}\in\Gamma}
\left[
\sum_{\mu=0}^1\frac{1}{2}
\left(
\phi({\bm r}+\hat\mu)-\phi({\bm r})
\right)^2
+
V(\phi({\bm r}))
\right].
\ee
Here $\hat\mu$ is the unit vector for the $\mu$ direction.
In the potential term $V$, the mass term and the self-interaction term are included, but here we do not specify them since
such detailed information is irrelevant in the following discussion.
Here we only assume that the potential $V$ is bounded from below.

\begin{figure}[t!]
\begin{center}
\begin{tabular}{c}
\includegraphics[width=5cm,pagebox=cropbox,clip]{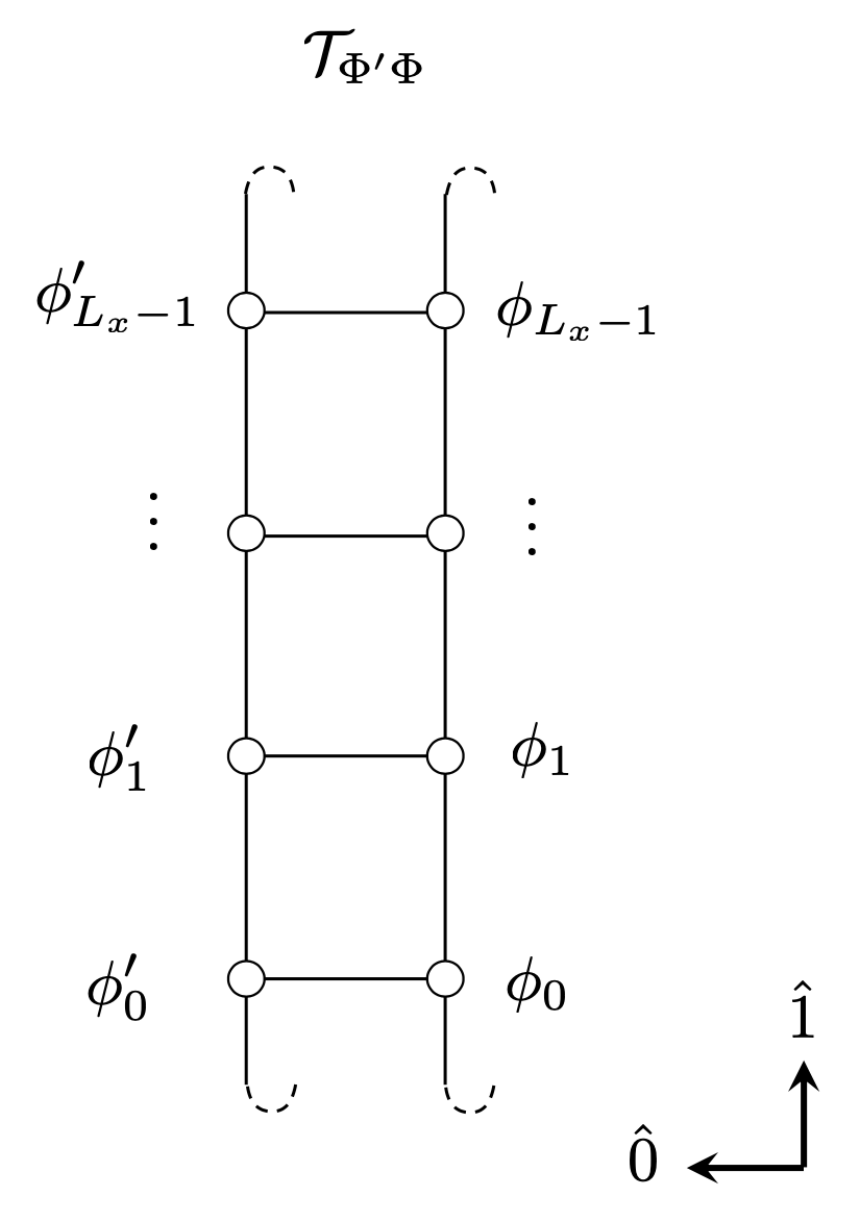}
\end{tabular}
\end{center}
\caption{
An image of the transfer matrix for the lattice field theory.
$\phi^\prime_x=\phi(t+1,x)$ and $\phi_x=\phi(t,x)$ with $x=0,1,2,\ldots,L_x-1$.
(Note that this figure is not a tensor network representation)
}
\label{fig:transfer_matrix}
\end{figure}

The partition function can be represented by
\be
Z=
{\rm Tr}
\left[
{\cal T}^{L_\tau}
\right]
\ee
where the transfer matrix ${\cal T}$ is given by \cite{montvay1994quantum}
\bea
{\cal T}[\Phi^\prime,\Phi]
&=&
\exp
\left[
-
\sum_{x=0}^{L_x-1}
\frac{1}{2}
\left(
\phi(t+1,x)-\phi(t,x)
\right)^2
-
\frac{1}{2}
L[\Phi^\prime]
-
\frac{1}{2}
L[\Phi]
\right],
\label{eqn:transfer_matrix}
\\
L[\Phi]
&=&
\sum_{x=0}^{L_x-1}
\left[
\frac{1}{2}
\left(
\phi(t,x+1)-\phi(t,x)
\right)^2
+
V(\phi(t,x))
\right],
\eea
with the field configurations on the Euclidean time slice at $t+1$ and $t$,
\begin{alignat}{4}
&\Phi^\prime&=&\{\phi(t+1&,&x)&|&x=0,1,2,\hdots,L_x-1\},
\\
&\Phi&=&\{\phi(t&,&x)&|&x=0,1,2,\hdots,L_x-1\}.
\label{eqn:integrated_phi}
\end{alignat}
See Fig.~\ref{fig:transfer_matrix} for a pictorial expression of the transfer matrix.
The transfer matrix for the continuous fields
is an integration kernel operator but in the following we treat it as if it were a usual matrix,
that is, $\Phi$ is treated as an integer-valued index just for notational simplicity:
\be
{\cal T}[\Phi^\prime,\Phi]
=
\langle\Phi^\prime|\hat{\cal T}|\Phi\rangle
=
{\cal T}_{\Phi^\prime\Phi}.
\ee

Since ${\cal T}$ is Hermitian and positive definite in the model of interest, it has the following eigenvalue decomposition (EVD)
\be
{\cal T}_{\Phi^\prime \Phi}
=
\sum_{a=0}^\infty U_{\Phi^\prime a}\lambda_a (U^\dag)_{a\Phi}
\,\,
\Longleftrightarrow
\,\,
\hat {\cal T}|a\rangle=\lambda_a|a\rangle
\label{eqn:T_EVD}
\ee
where $U_{\Phi a}=\langle\Phi|a\rangle$ is the unitary matrix\footnote{
For a continuous variable $\Phi$, the unitary matrix $U_{\Phi a}$ should be replaced by a set of orthonormal eigenfunctions $u_{a}(\Phi)$
that satisfy the orthonormal condition
$
\int d\Phi \,\,u_a(\Phi) u_b^\ast(\Phi)=\delta_{ab}$ 
and the completeness
$
\sum_a u_a(\Phi) u_a^\ast(\Phi^\prime)=\delta(\Phi-\Phi^\prime).
$
}
composed from the eigenstates and $\lambda_a$ are the eigenvalues and assumed to be arranged in descending order, $\lambda_0\ge\lambda_1\ge\hdots$.
The energy gaps defined in Eq. (\ref{eqn:energy_gap}) may be estimated from the transfer matrix spectrum $\lambda_a$
up to lattice cutoff effects,
\be
\omega_a
=
\ln\left(\frac{\lambda_0}{\lambda_a}\right)
\hspace{10mm}
\mbox{ for } a=1,2,3,\ldots.
\ee

In this way, the diagonalization of the transfer matrix tells us the energy eigenvalues and eigenstates
but quantum numbers of each eigenstate are not \textit{a priori} known; that is,
at this stage the correspondence between $(n,q)$ in Eq. (\ref{eqn:energy_eigen_equation}) and $a$ is not clear.
To identify the quantum numbers of the eigenstates, an additional procedure is required as follows.
First, we have to prepare matrix elements of an interpolating operator $\hat{\cal O}_q$ between the energy eigenstates,
\be
B_{ba}:=
\langle b |\hat {\cal O}_q|a\rangle
=
(U^\dag {\cal O}_q U)_{ba},
\label{eqn:matrix_element}
\ee
where $U$ is the unitary matrix given in Eq. (\ref{eqn:T_EVD}) and ${\cal O}_q$ is a field representation of the interpolating operator,
\be
({\cal O}_{q})_{\Phi^\prime\Phi}
=
\langle\Phi^\prime|\hat{\cal O}_q|\Phi\rangle.
\ee

In order to explain how the matrix element is used to determine the quantum number of the eigenstates,
let us derive a selection rule for a given symmetry.
First, we consider the continuous symmetry case.
Let $\hat Q$ be a conserved charge associated with the symmetry, and
it satisfies $[\hat Q,\hat {\cal T}]=0$.
The associated quantum number (or charge) for some operator $\hat X$ is denoted by $q_X$,
\be
[\hat Q,\hat X]=q_X\hat X.
\label{eqn:QX}
\ee
Assuming that the ground state has no charge $\hat Q|\Omega\rangle=0$,
Eq. (\ref{eqn:QX}) tells us that $\hat X|\Omega\rangle$ is an eigenstate of $\hat Q$ whose
quantum number is $q_X$.
By sandwiching Eq. (\ref{eqn:QX}) between $\langle a|$ and $|b\rangle$,
one obtains a relationship between the charges and the matrix elements,
\be
(q_b-q_a-q_X)
\langle b|\hat X|a\rangle=0
\label{eqn:relation_continuum}
\ee
where the quantum number of $|a\rangle$ is assumed to be represented as $q_a$:
\be
\hat Q|a\rangle=q_a|a\rangle.
\ee
From Eq. (\ref{eqn:relation_continuum}), we see a selection rule for the continuous symmetry :
\begin{alignat}{3}
\mbox{for } &\langle b|\hat X|a\rangle&\neq&0 \,\,\,\Longrightarrow\,\,\, &q_b-q_a-q_X&=0.
\non
\end{alignat}
The selection rule can be used to identify the quantum number of the transfer matrix eigenstates.
For example, when we consider Eq. (\ref{eqn:relation_continuum})
with setting $a=0$, which means the ground state\footnote{In the finite spatial volume, the spontaneous symmetry breaking
does not occur thus we can assume that the ground state is not degenerated.}
and its quantum number is zero $q_a=0$, 
we can say that
if the matrix element is $\langle b|\hat X|\Omega\rangle\neq0$ then the quantum number of $|b\rangle$ is
shown to be $q_b=q_X$.
In this way, the matrix elements tell us the quantum number of the eigenstates.

A similar argument holds for the discrete symmetry.
Let $\hat D$ be a discrete transformation operator and we assume that the discrete transformation
for an operator $\hat X$ is given by
\be
\hat D \hat X \hat D^{-1}=q_X \hat X
\label{eqn:DX}
\ee
where we call
$q_X=\pm1$
charge of $\hat X$ for the discrete symmetry.
The ground state is assumed to have the unit charge $\hat D|\Omega\rangle=+|\Omega\rangle$.
Then by using Eq. (\ref{eqn:DX}) one can show that
$\hat X|\Omega\rangle$ is an eigenstate of $\hat D$ whose eigenvalue is $q_X$,
namely, $\hat D\hat X|\Omega\rangle=q_X\hat X|\Omega\rangle$.
From Eq. (\ref{eqn:DX}), one can derive a relationship between the charges and the matrix elements,
\be
\langle b|\hat X|a\rangle
=
\langle b|
\hat D^{-1}\hat D
\hat X
\hat D^{-1}\hat D
|a\rangle
=
q_bq_aq_X
\langle b|\hat X|a\rangle,
\label{eqn:relation_discrete}
\ee
for the eigenstate $|a\rangle$ whose charge is denoted by $q_a$,
\be
\hat D|a\rangle=q_a|a\rangle.
\ee
From Eq. (\ref{eqn:relation_discrete}), we read a selection rule for the discrete symmetry :
\begin{alignat}{3}
\mbox{for }  & \langle b|\hat X|a\rangle&\neq0 &\,\,\,\Longrightarrow\,\,\, &q_bq_aq_X&=1.
\non
\end{alignat}

As seen in this subsection, the transfer matrix formalism, in principle,
cannot only obtain the spectrum of a system but also determine the quantum number of the eigenstates.
There is, however, a practical difficulty for the formalism.
For large lattice volume, the dimension of the transfer matrix becomes extremely large
and the numerical computation cannot be done.
Thus, one has to rely on approximation methods.

\subsection{How to compute energy spectrum and matrix elements}
\label{sec:how_to}
As mentioned at the end of the previous subsection,
although the transfer matrix formalism is rather theoretically apparent,
it is practically very difficult to numerically make the transfer matrix itself for the lattice field theory
since the dimensionality of the transfer matrix becomes extremely large.
One way to avoid such a problem is to approximately
make the transfer matrix by using the tensor network method.
In this case, the dimensionality of the transfer matrix can be drastically reduced and one can numerically treat it as we will see.

\begin{figure}[t!]
\begin{center}
\begin{tabular}{c}
\includegraphics[width=11cm,pagebox=cropbox,clip]{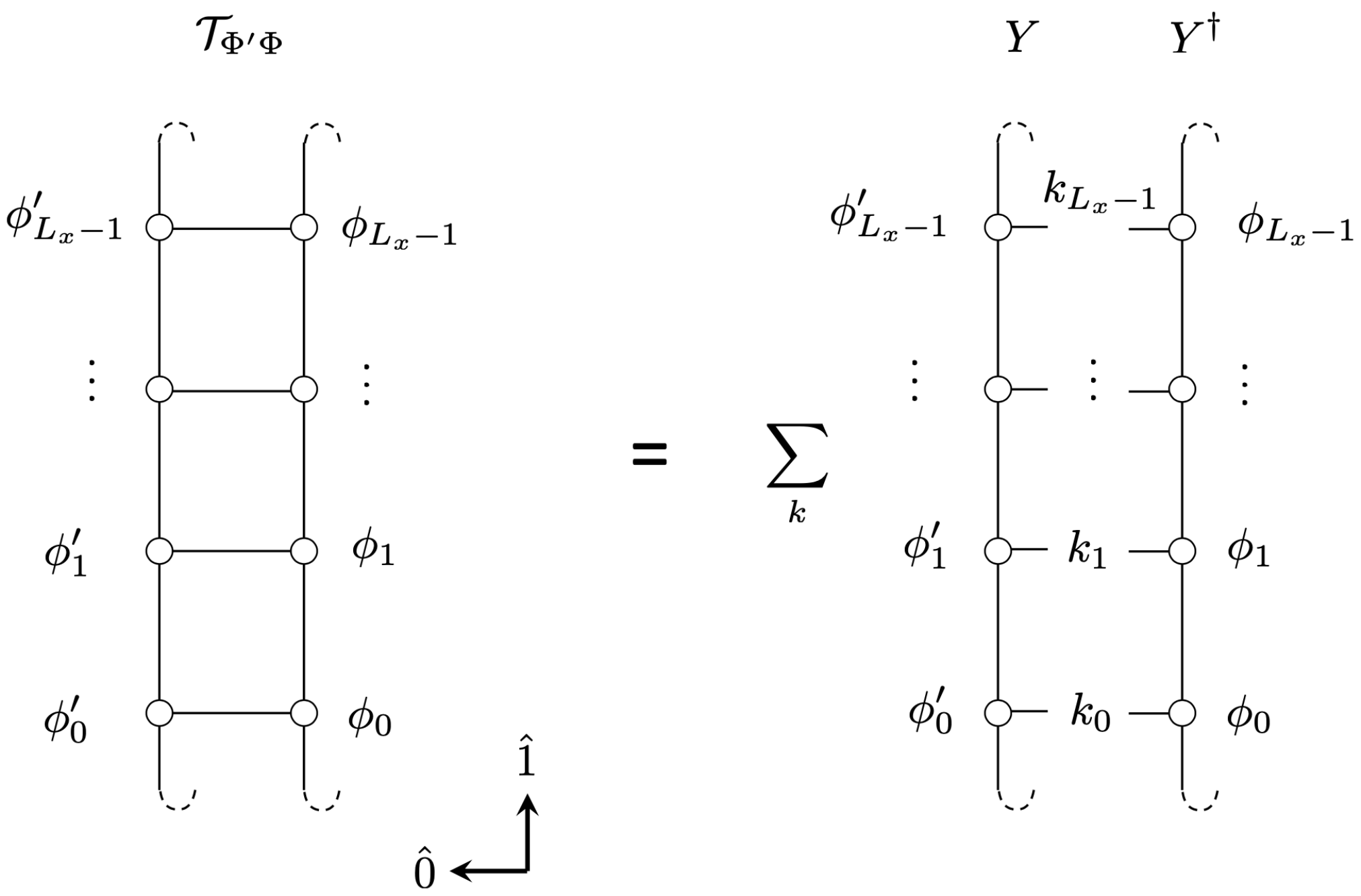}
\end{tabular}
\end{center}
\caption{
Graphical expression of the decomposition of the transfer matrix in terms of $Y$ in Eq. (\ref{eqn:TM_YY}).
(Note that this figure is not a tensor network representation)
}
\label{fig:transfer_matrix_YY}
\end{figure}

The starting point is the definition of the transfer matrix in Eq. (\ref{eqn:transfer_matrix}) but here
we rewrite it for convenience in the following discussion :
\bea
{\cal T}_{\Phi^\prime\Phi}
&=&
\left(
\prod_{x=0}^{L_x-1}
\exp\left[
-\frac{1}{2}(\phi_x^\prime-\phi_x)^2
-\frac{1}{4}V(\phi^\prime_x)
-\frac{1}{4}V(\phi_x)
\right]
\right)
\non\\
&\times&
\left(
\prod_{x=0}^{L_x-1}
\exp\left[-\frac{1}{4}(\phi_{x+1}^\prime-\phi_x^\prime)^2
-\frac{1}{8}V(\phi^\prime_{x+1})
-\frac{1}{8}V(\phi^\prime_x)
\right]
\right)
\non\\
&\times&
\left(
\prod_{x=0}^{L_x-1}
\exp\left[-\frac{1}{4}(\phi_{x+1}-\phi_x)^2
-\frac{1}{8}V(\phi_{x+1})
-\frac{1}{8}V(\phi_x)
\right]
\right),
\label{eqn:TM_details}
\eea
where $\phi^\prime_x=\phi(t+1,x)$ and $\phi_x=\phi(t,x)$ with $x=0,1,2,\hdots,L_x-1$.
The first term represents a hopping for the time direction.
The second and third terms are for the space direction at the time slice $t+1$ and $t$, respectively.
If we apply the eigenvalue decomposition (EVD) to
each local Boltzmann weight for the time hopping
in Eq. (\ref{eqn:TM_details}),
\be
\exp\left[
-\frac{1}{2}(\phi_x^\prime-\phi_x)^2
-\frac{1}{4}V(\phi^\prime_x)
-\frac{1}{4}V(\phi_x)
\right]
=
\sum_{k_x=0}^\infty
u_{\phi^\prime_x{k_x}}
\sigma_{k_x}
(u^\dag)_{k_x\phi_x},
\label{eqn:hopping1}
\ee
then we can decompose the transfer matrix as follows:
\be
{\cal T}_{\Phi^\prime\Phi}
=
\sum_k Y_{\Phi^\prime k} Y^\dag_{k\Phi},
\label{eqn:TM_YY}
\ee
with
\bea
Y_{\Phi k}
&=&
\left(
\prod_{x=0}^{L_x-1}
\sum_{k_x=0}^\infty
u_{\phi_x{k_x}}
\sqrt{\sigma_{k_x}}
\right)
\non\\&\times&
\left(
\prod_{x=0}^{L_x-1}
\exp\left[-\frac{1}{4}(\phi_{x+1}-\phi_x)^2
-\frac{1}{8}V(\phi_{x+1})
-\frac{1}{8}V(\phi_x)
\right]
\right),
\eea
where we have defined the integrated index,
$k=(k_0,k_1,k_2,\hdots,k_{L_x-1})$.
An image of the decomposition in Eq. (\ref{eqn:TM_YY}) is shown in Fig.~\ref{fig:transfer_matrix_YY}.

\begin{figure}[t!]
\begin{center}
\begin{tabular}{c}
\includegraphics[width=11cm,pagebox=cropbox,clip]{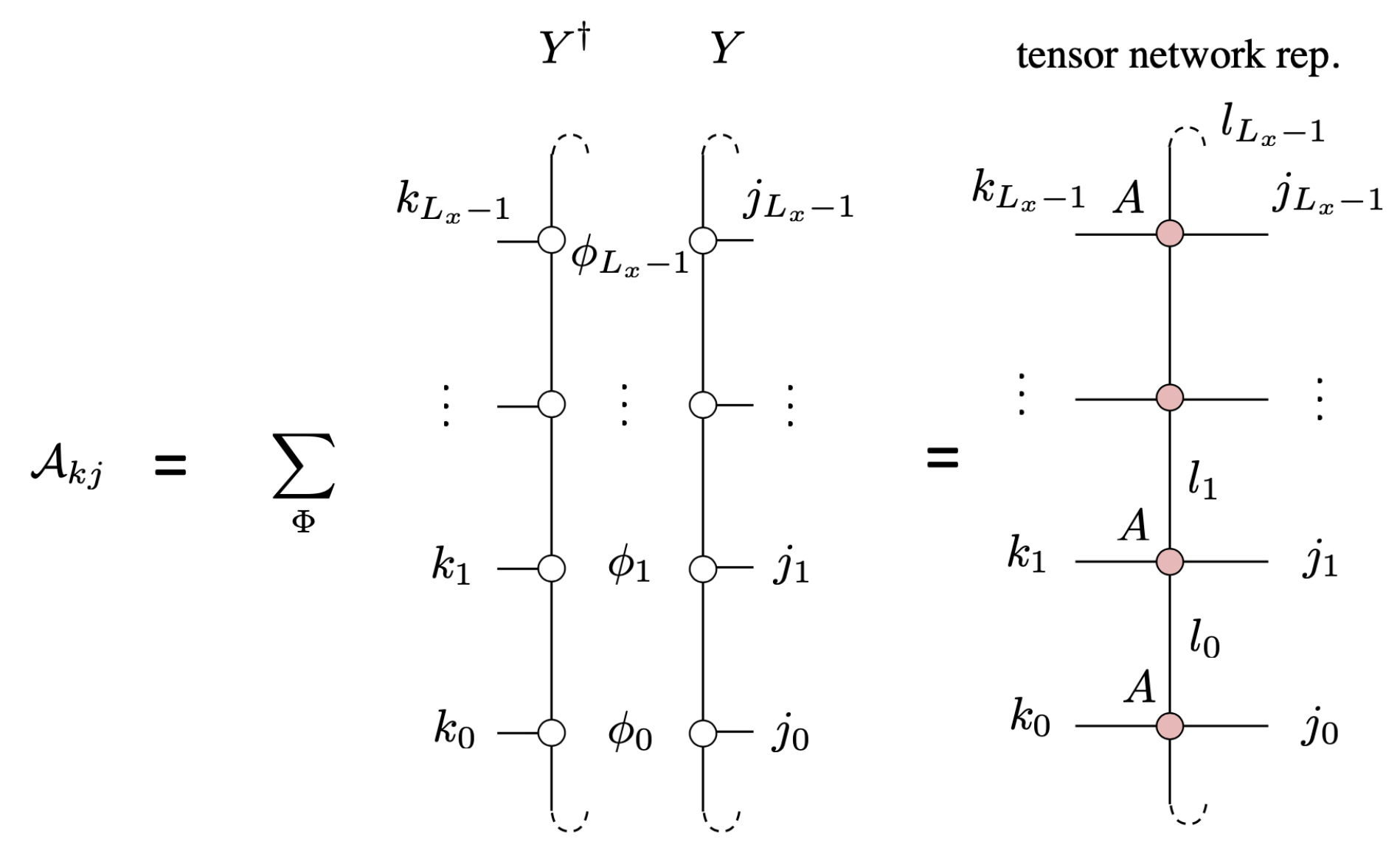}
\end{tabular}
\end{center}
\caption{
The definition of
${\cal A}$
in Eq. (\ref{eqn:tildeT_YY}) and
its tensor network representation in terms of the local tensor $A$
in Eq. (\ref{eqn:initial_tensor}).
}
\label{fig:tilde_transfer_matrix_YY}
\end{figure}

Substituting the transfer matrix in Eq. (\ref{eqn:TM_YY}) into the partition function, we obtain
\be
Z
=
{\rm Tr}_{\Phi}
\left[
{\cal T}^{L_\tau}
\right]
=
{\rm Tr}_{\Phi}
\left[
(Y Y^\dag)^{L_\tau}
\right]
=
{\rm Tr}_{k}
\left[
(Y^\dag Y)^{L_\tau}
\right]
=
{\rm Tr}_{k}
\left[
{\cal A}^{L_\tau}
\right],
\ee
where in the last equal we have defined
\be
{\cal A}:=Y^\dag Y.
\label{eqn:tildeT_YY}
\ee
Note that the ordering of $Y$ and $Y^\dag$ is different in ${\cal T}$ and
${\cal A}$.
As shown in Fig.~\ref{fig:tilde_transfer_matrix_YY}, actually
${\cal A}$
can be simply expressed by a tensor network representation
\bea
{\cal A}_{kj}
&=&
\sum_{\Phi}
(Y^\dag)_{k\Phi}Y_{\Phi j}
\non\\
&=&
\sum_{\Phi}
\left(
\prod_{x=0}^{L_x-1}
\sum_{k_x=0}^\infty
(u^\dag)_{k_x\phi_x}
\sqrt{\sigma_{k_x}}
\right)
\left(
\prod_{x=0}^{L_x-1}
\sum_{j_x=0}^\infty
u_{\phi_x{j_x}}
\sqrt{\sigma_{j_x}}
\right)
\non\\
&\times&
\left(
\prod_{x=0}^{L_x-1}
\exp\left[-\frac{1}{2}(\phi_{x+1}-\phi_x)^2
-\frac{1}{4}V(\phi_{x+1})
-\frac{1}{4}V(\phi_x)
\right]
\right)
\non\\
&=&
\prod_{x=0}^{L_x-1}
\sum_{l_x=0}^\infty
A_{k_xl_xj_xl_{x-1}}
\label{eqn:tildeT_A}
\eea
with a rank-four tensor $A$
\be
A_{k_xl_xj_xl_{x-1}}
:=
\sqrt{\sigma_{k_x}\sigma_{l_x}\sigma_{j_x}\sigma_{l_{x-1}}}
\sum_{\phi_x}
(u^\dag)_{k_x\phi_x}
(u^\dag)_{l_x\phi_x}
u_{\phi_xj_x}
u_{\phi_xl_{x-1}}.
\label{eqn:initial_tensor}
\ee
When deriving the initial tensor $A$, we have applied the EVD to the spatial hopping terms in Eq. (\ref{eqn:tildeT_A}),
\be
\exp\left[
-\frac{1}{2}(\phi_{x+1}-\phi_x)^2
-\frac{1}{4}V(\phi_{x+1})
-\frac{1}{4}V(\phi_x)
\right]
=
\sum_{l_x=0}^\infty
u_{\phi_{x+1}{l_x}}
\sigma_{l_x}
(u^\dag)_{l_x\phi_x},
\label{eqn:hopping2}
\ee
for $x\in\{0,1,2,\hdots,L_{x}-1\}$.
 
Note that the singular value decomposition of $Y$ is given by
\be
Y=U\sqrt{\lambda} W^\dag
\label{eqn:Y_SVD}
\ee
where $U$ and $\lambda$ are the same as those of the transfer matrix ${\cal T}$ in Eq. (\ref{eqn:T_EVD}).
On the other hand, the EVD for
${\cal A}$
is given by
\be
{\cal A}
=
Y^\dag Y
=
W\lambda W^\dag,
\label{eqn:tildeTM_EVD}
\ee
thus
${\cal A}$
has the same eigenvalues as those of ${\cal T}$.

\begin{figure}[t!]
\begin{center}
\begin{tabular}{c}
\includegraphics[width=14cm,pagebox=cropbox,clip]{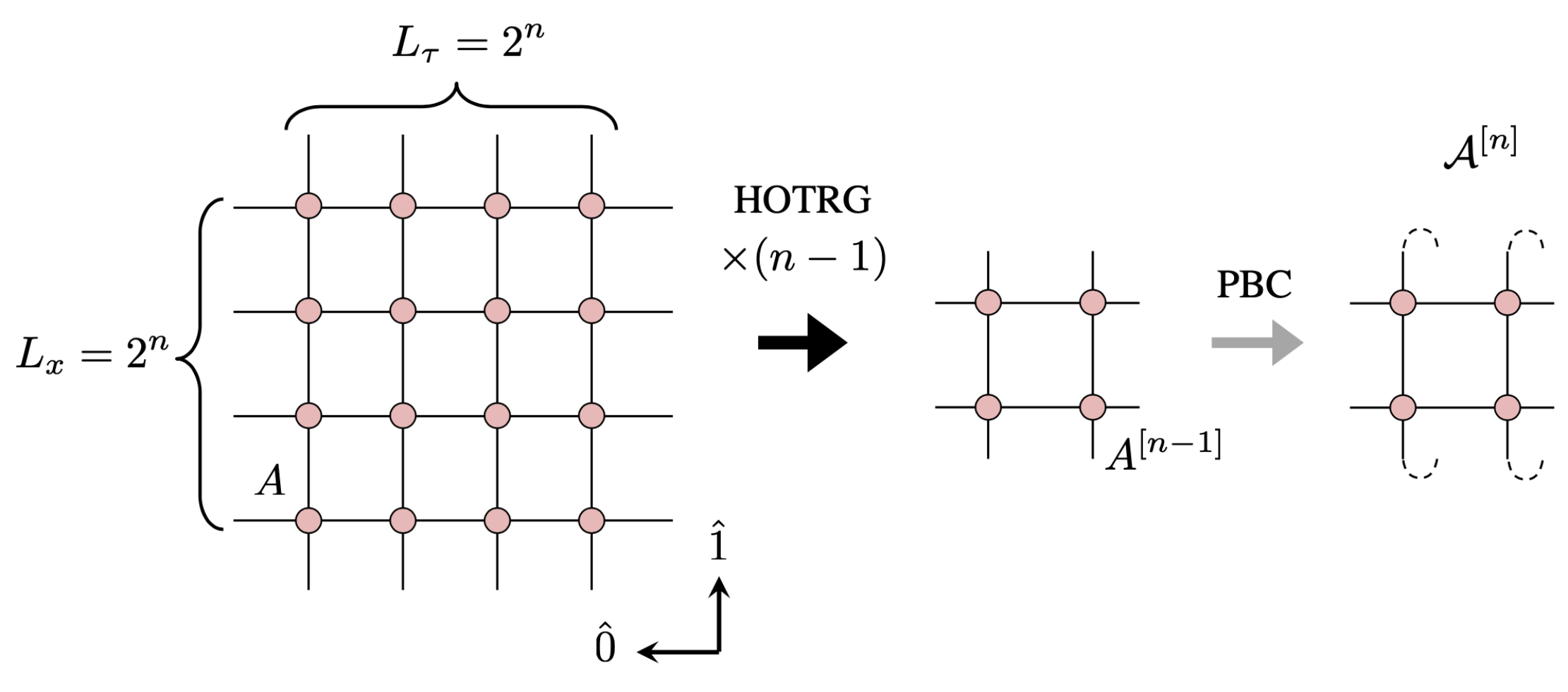}
\end{tabular}
\end{center}
\caption{
The black arrow shows the coarse-graining procedure of the tensor network.
The procedure is done until there are only four ${A}^{[n-1]}$s.
The gray arrow shows
the direct contraction of the last four tensors to obtain the approximation of
${\cal A}^{L_\tau}\approx{\cal A}^{[n]}$.
Here the PBC is imposed on the space direction.
}
\label{fig:coarse-graining}
\end{figure}

From here let us explain how to numerically compute the spectrum
by using the tensor network method \cite{PhysRevB.107.205123}.
First, we coarse grain\footnote{
Before coarse graining, we have to truncate the summation in the initial tensor network and set a bond dimension for the initial tensor.}
 the tensor network consisting of the initial tensor $A$ in Eq. (\ref{eqn:initial_tensor})
on the square lattice
with $L_\tau=L_x=2^n$ by using HOTRG \cite{PhysRevB.86.045139} as shown in Fig.~\ref{fig:coarse-graining},
\be
A=A^{[0]}
\stackrel{{\rm HOTRG}}{\longrightarrow} A^{[1]}
\stackrel{{\rm HOTRG}}{\longrightarrow} A^{[2]}
\stackrel{{\rm HOTRG}}{\longrightarrow} \hdots.
\ee
and then after $n-1$ iterations, one arrives at four renormalized tensors $A^{[n-1]}$.
Subsequently we perform the direct contraction of those tensors taking into account the periodic boundary condition on the spatial direction,
and we obtain a numerical approximation of a power of
${\cal A}$ as follows:
\begin{equation}
{\cal A}^{L_\tau}
\approx
{\mathcal{A}}^{[n]}_{k_1k_2,j_1j_2}
=
\sum_{l_1,l_2,m_1,m_2,n_1,n_2}
A^{[n-1]}_{k_1l_1n_1l_2}A^{[n-1]}_{k_2l_2n_2l_1}
A^{[n-1]}_{n_1m_1j_1m_2}A^{[n-1]}_{n_2m_2j_2m_1}.
\label{eqn:tildeT_n}
\end{equation}
Subsequently, we diagonalize\footnotemark\footnotetext{
One may define
${\cal A}^{[n]}$
from a single coarse-grained tensor $A^{[n]}$ after $n$ steps instead of Eq. (\ref{eqn:tildeT_n}).
In this case, however, the numerical matrix is not guaranteed to be positive definite and, furthermore, we find that the accuracy of the spectrum is not so good.
On the other hand, thanks to $2\times2$ structure, the matrix
${\cal A}^{[n]}$
in Eq. (\ref{eqn:tildeT_n}) is manifestly positive definite
and, furthermore, the accuracy of the spectrum is much better than the former case.
}
$\mathcal{A}^{[n]}$
as follows:
\begin{equation}
\mathcal{A}^{[n]}
=W^{[n]}\lambda^{[n]}W^{[n]\dagger}
\label{eqn:tildeT_EVD}
\end{equation}
where $W^{[n]}$ is a unitary matrix containing numerical eigenvectors, and $\lambda^{[n]}$ is the eigenvalues.
Note that the original lattice size in the time direction of
${\cal A}^{[n]}$
is $L_{\tau}(=2^{n})$, thus
the tensor network estimation of the transfer matrix eigenvalue is given by
\begin{eqnarray}\label{lambdaapx}
\lambda_a\approx\left(\lambda_a^{[n]}\right)^{1/L_{\tau}}
\end{eqnarray}
then the energy gap is estimated as
\begin{equation}\label{egaptn}
\omega_a\approx\frac{1}{L_{\tau}}\log\left(\frac{\lambda_0^{[n]}}{\lambda_a^{[n]}}\right)
=:
\omega_a^{[{\rm hotrg}]}
\,\,\,\,
\mbox{ for } a=1,2,3,\hdots.
\end{equation}

\begin{figure}[t!]
\begin{center}
\begin{tabular}{c}
\includegraphics[width=12cm,pagebox=cropbox,clip]{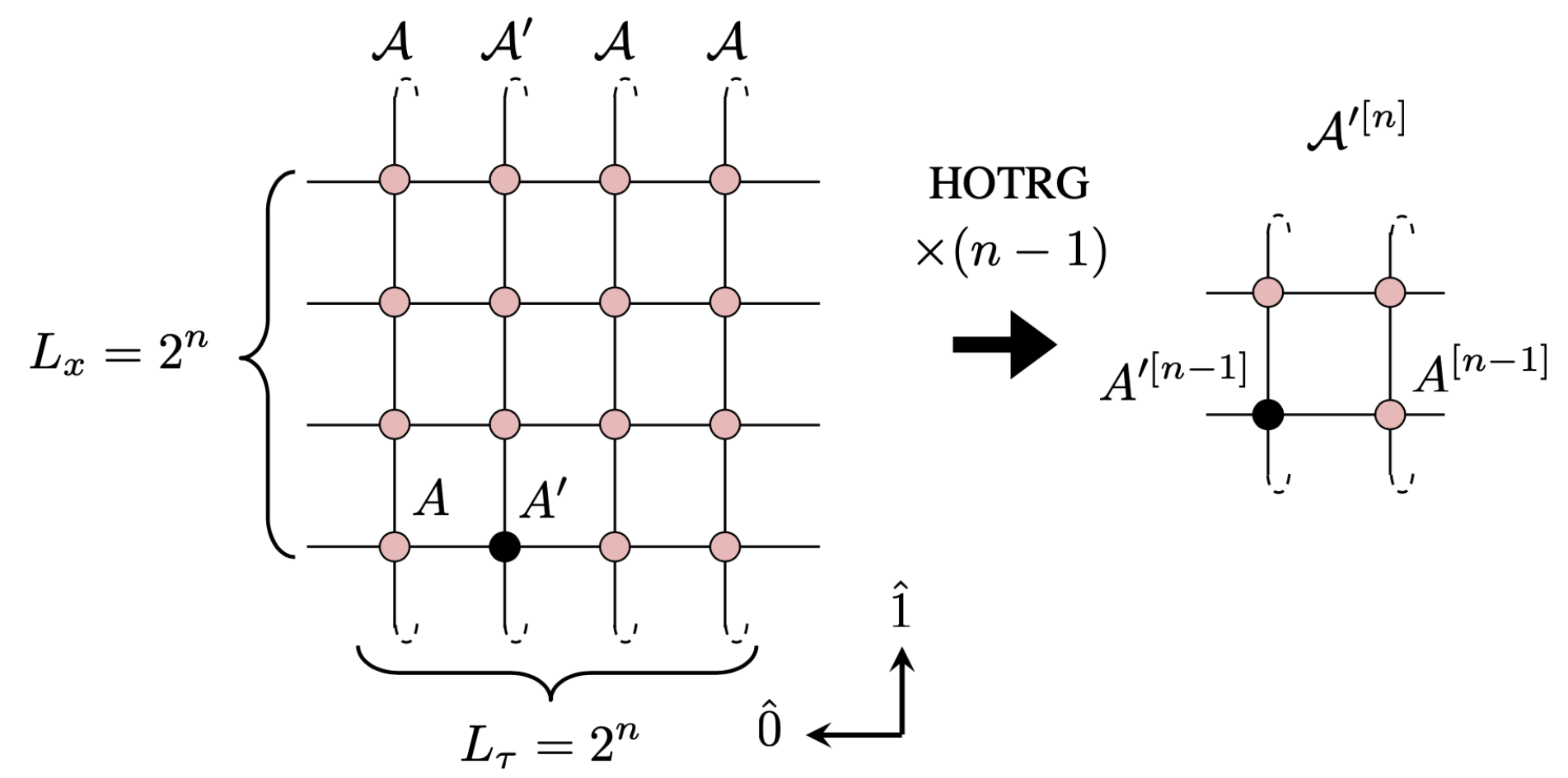}
\end{tabular}
\end{center}
\caption{
The coarse-graining procedure of the impurity tensor network ${\cal A}^{m-1}{\cal A}^\prime {\cal A}^m$
with the single field operator ${\cal O}_q=\phi_x$.
The procedure is done until there are four tensors.
}
\label{fig:coarse-graining_impurity}
\end{figure}

Next, let us see how to compute the matrix elements by the tensor network method.
For that purpose, first we rewrite the matrix elements in Eq. (\ref{eqn:matrix_element}) 
in terms of the tensor network related quantities.
For an integer $m=L_{\tau}/2$ \footnote{The reason why we take $m=L_\tau/2$ will be explained around Eq. (\ref{eqn:Ap}). }
(assuming that $L_\tau$ is even number), the matrix elements 
may be expressed as follows: 
\bea
B_{ba}
&=&
\langle b|\hat {\cal O}_q|a\rangle
=
(U^\dag {\cal O}_q U)_{ba}
\non\\&=&
(
U^\dag {\cal T}^{-m} {\cal T}^{m}
{\cal O}_q {\cal T}^{m+1} {\cal T}^{-(m+1)} U
)_{ba}
\,\,\,(\mbox{using }{\cal T}{\cal T}^{-1}=1)
\non\\&=&
(
U^\dag (U\lambda U^\dag)^{-m} (YY^\dag)^{m}
{\cal O}_q (YY^\dag)^{m+1}(U\lambda U^\dag)^{-(m+1)} 
U
)_{ba}
\,\,\,
[\mbox{using Eqs. (\ref{eqn:T_EVD}) and (\ref{eqn:TM_YY})}]
\non\\&=&
(
\lambda^{-m} U^\dag Y(Y^\dag Y)^{m-1}Y^\dag
{\cal O}_q 
Y (Y^\dag Y)^{m}Y^\dag U \lambda^{-(m+1)}
)_{ba}
\non\\&=&
(
\lambda^{-m} U^\dag Y
{\cal A}^{m-1}
{\cal A}^\prime
{\cal A}^{m}Y^\dag U \lambda^{-(m+1)}
)_{ba}
\,\,\,
[\mbox{using Eq. (\ref{eqn:tildeT_YY}) and } {\cal A}^\prime:=Y^\dag{\cal O}_qY]
\non\\&=&
(
\lambda^{-m+1/2} W^\dag
{\cal A}^{m-1}
{\cal A}^\prime
{\cal A}^{m}W \lambda^{-m-1/2}
)_{ba}
\,\,\,
[\mbox{using Eq. (\ref{eqn:Y_SVD})}],
\label{eqn:B}
\eea
where we have defined an impurity version of ${\cal A}$, that is,
${\cal A}^\prime:=Y^\dag{\cal O}_qY$
and an impurity tensor network ${\cal A}^{m-1}{\cal A}^\prime {\cal A}^m$ as shown in Fig.~\ref{fig:coarse-graining_impurity}.
Here we assume that the lattice size for the impurity tensor network is the same
as that of pure tensor network $L_\tau=L_x=2^n$.
If we consider a single field 
${\cal O}_q=\phi_{x}$ at a lattice site $x$, then the associated impurity tensor $A^\prime$ is given by
\be
A_{k_xl_xj_xl_{x-1}}^\prime
:=
\sqrt{\sigma_{k_x}\sigma_{l_x}\sigma_{j_x}\sigma_{l_{x-1}}}
\sum_{\phi_x}
\phi_x
(u^\dag)_{k_x\phi_x}
(u^\dag)_{l_x\phi_x}
u_{\phi_xj_x}
u_{\phi_xl_{x-1}}.
\label{eqn:impurity_tensor}
\ee
In this way, one can represent the matrix elements in terms of the impurity tensor network ${\cal A}^{m-1}{\cal A}^\prime {\cal A}^m$, $W$ and $\lambda$ that are obtained from the EVD of ${\cal A}$ as in Eq. (\ref{eqn:tildeTM_EVD}).

In order to numerically evaluate the impurity tensor network ${\cal A}^{m-1}{\cal A}^\prime{\cal A}^m$,
we apply the coarse-graining procedure to this network using
the same isometries as in the pure tensor network coarse-graining steps shown in Fig.~\ref{fig:coarse-graining}
until there are four tensors.
We denote the coarse-grained impurity tensor network ($2\times2$ network in Fig.~\ref{fig:coarse-graining_impurity}) as
${\cal A}^{\prime[n]}$,
\be
{\cal A}^{m-1}{\cal A}^\prime {\cal A}^m
\approx
{\cal A}^{\prime[n]}.
\label{eqn:Ap}
\ee
Some readers may wonder why we choose $m=L_\tau/2$ but not some small value or just using a one-time slice object that
means 
cheap computational cost.
In the latter case, however, we find that an accuracy of the spectrum and the evaluation of the impurity tensor network turns out to be worse (see Appendix \ref{sec:1d_TN} for more details of a spectroscopy using a one-time slice transfer matrix, that is, one-dimensional coarse-graining scheme).
On the other hand, for $m=L_\tau/2$ in Eq. (\ref{eqn:Ap}) that corresponds to the square impurity tensor network,
the coarse graining procedure is rather simple and we find that
the accuracy is reasonably maintained during the coarse-graining, thus we choose this value of $m$.
By using 
${\cal A}^{\prime[n]}$ in Eq. (\ref{eqn:Ap}), 
$W^{[n]}$ and $\lambda^{[n]}$ in Eq. (\ref{eqn:tildeT_EVD}),
the matrix elements in Eq. (\ref{eqn:B}) may be estimated by
\be\label{eqn:B_apx}
B_{ba}
\approx
\left(
\lambda^{[n]}
\right)^{-\frac{L_\tau-1}{2L_\tau}}
W^{[n]\dag}
{\cal A}^{\prime[n]}
W^{[n]}
\left(
\lambda^{[n]}
\right)^{-\frac{L_\tau+1}{2L_\tau}}
=: B_{ba}^{[{\rm hotrg}]}.
\ee

\section{NUMERICAL RESULTS}\label{sec:res}
In this section, 
we demonstrate our scheme by applying it to the $(1+1)$d Ising model with zero external magnetic field and the periodic boundary condition.
The details for the model, say its transfer matrix and tensor network representation are given in Appendix \ref{sec:TN_TM_2DIsing}. 
We will show that the scheme can produce the energy spectrum of the model
and the result will be compared with the exact spectrum \cite{PhysRev.76.1232} summarized in Appendix \ref{sec:exact_spectrum}. 
The matrix elements for the model with a single spin field or double fields are also computed to judge the quantum number of the eigenstates.
Furthermore, the wave function of the eigenstates and the scattering phase shift are computed as well.

\subsection{Energy spectrum}\label{sec:energy_spectrum}

\begin{figure}[t!]
\centering
\begin{subfigure}[b]{0.4\textwidth}
\centering
\includegraphics[width=8cm,height=6.5cm]{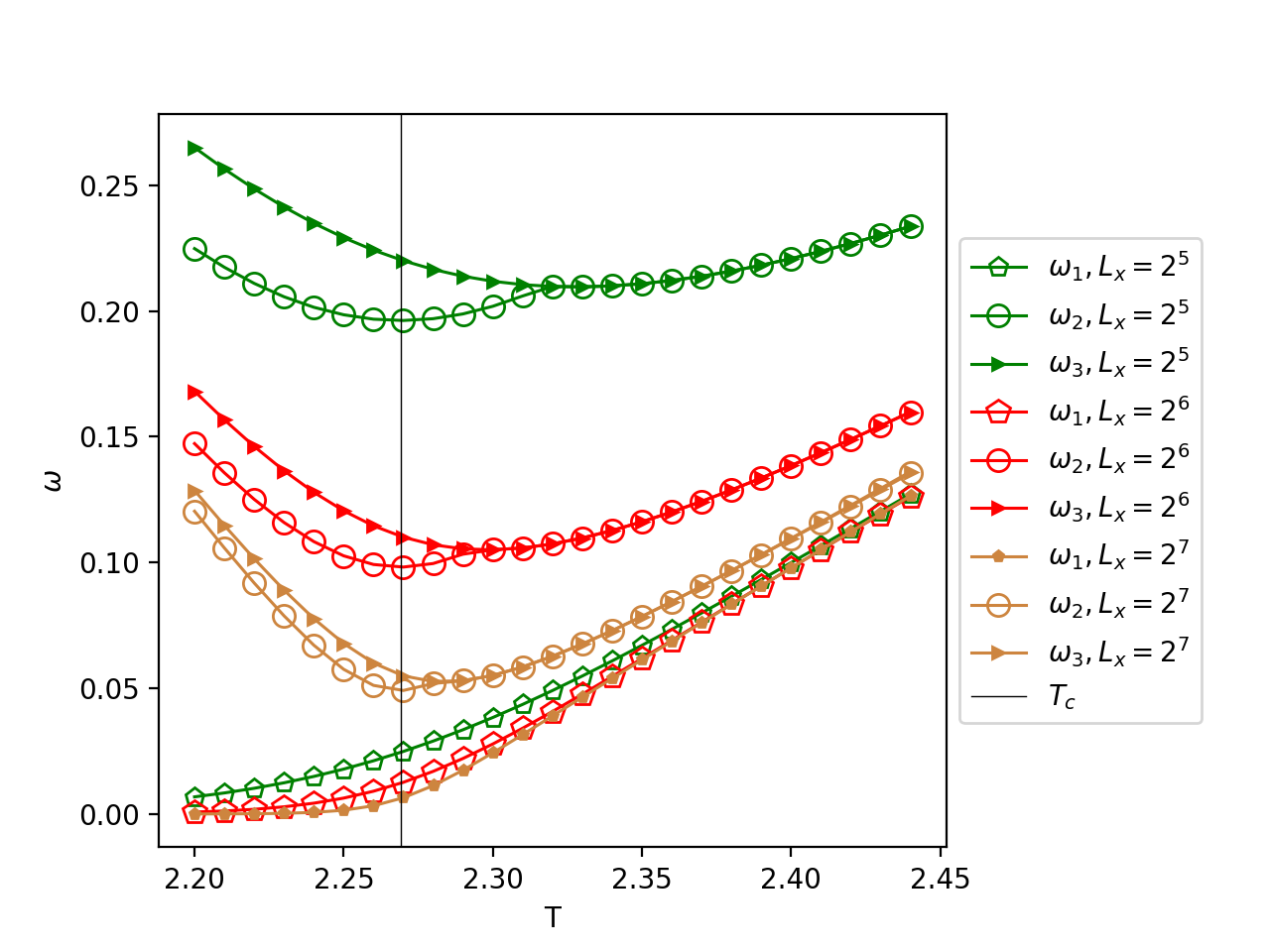}
\caption{}
\label{energy_s}
\end{subfigure}\hspace{20mm}
\begin{subfigure}[b]{0.4\textwidth}
\centering
\includegraphics[width=8cm,height=6.5cm]{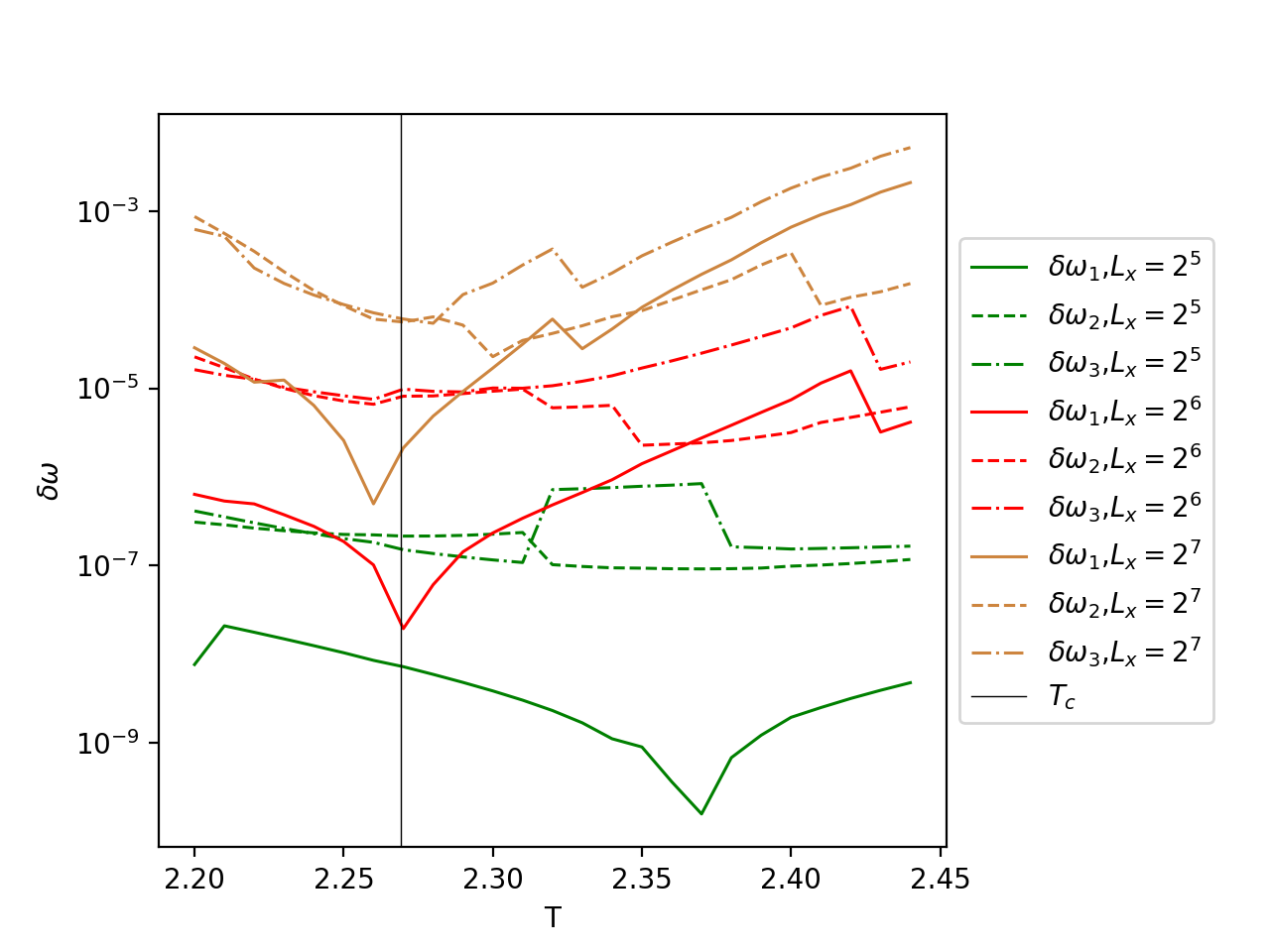}
\caption{}
\label{er_energy}
\end{subfigure}
\caption{(a) Three lowest energy gaps $\omega_a$ ($a=1,2,3$) for system size $L_{x}=2^5,2^6,2^7$ around the critical temperature $T_{\rm c}$ with $\chi=80$.  (b) The relative error of the energy gap $\delta\omega_a$ .
}
\label{energy_spectrum}
\end{figure}

\begin{figure}[t!]
\centering
\includegraphics[width=10cm,height=7.5cm]{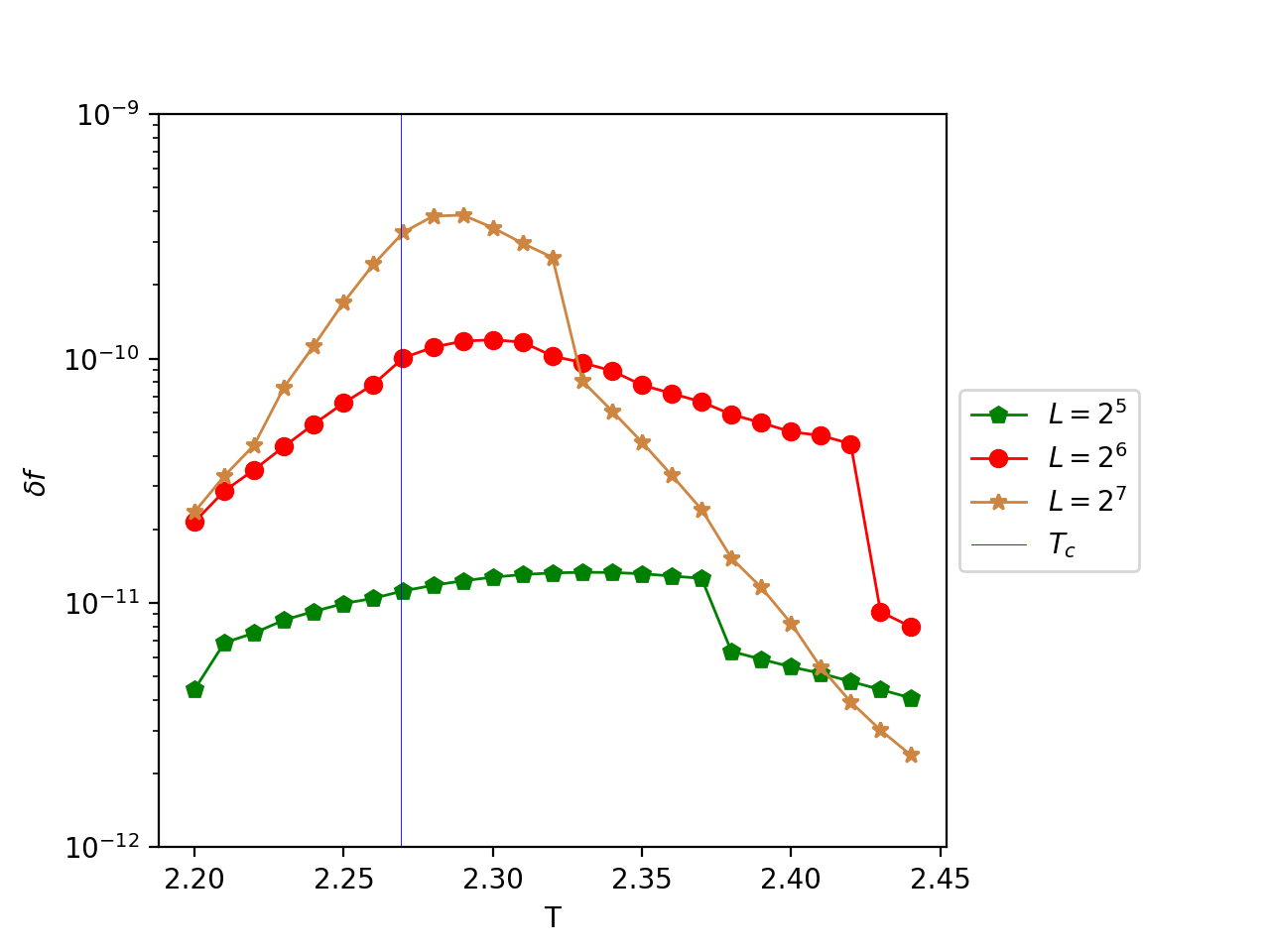}
\caption{The relative error of the free energy $\delta f$ at finite volume $V=L^2$ with $L=L_x=L_\tau$ computed by HOTRG with $\chi=80$. Around the critical point, $\delta f$ has a peak in contrast to $\delta \omega_1$ in Fig.~\ref{er_energy}.
}
\label{free_energy}
\end{figure}

\begin{figure}[t!]
\centering
\includegraphics[width=10cm,height=7.5cm]{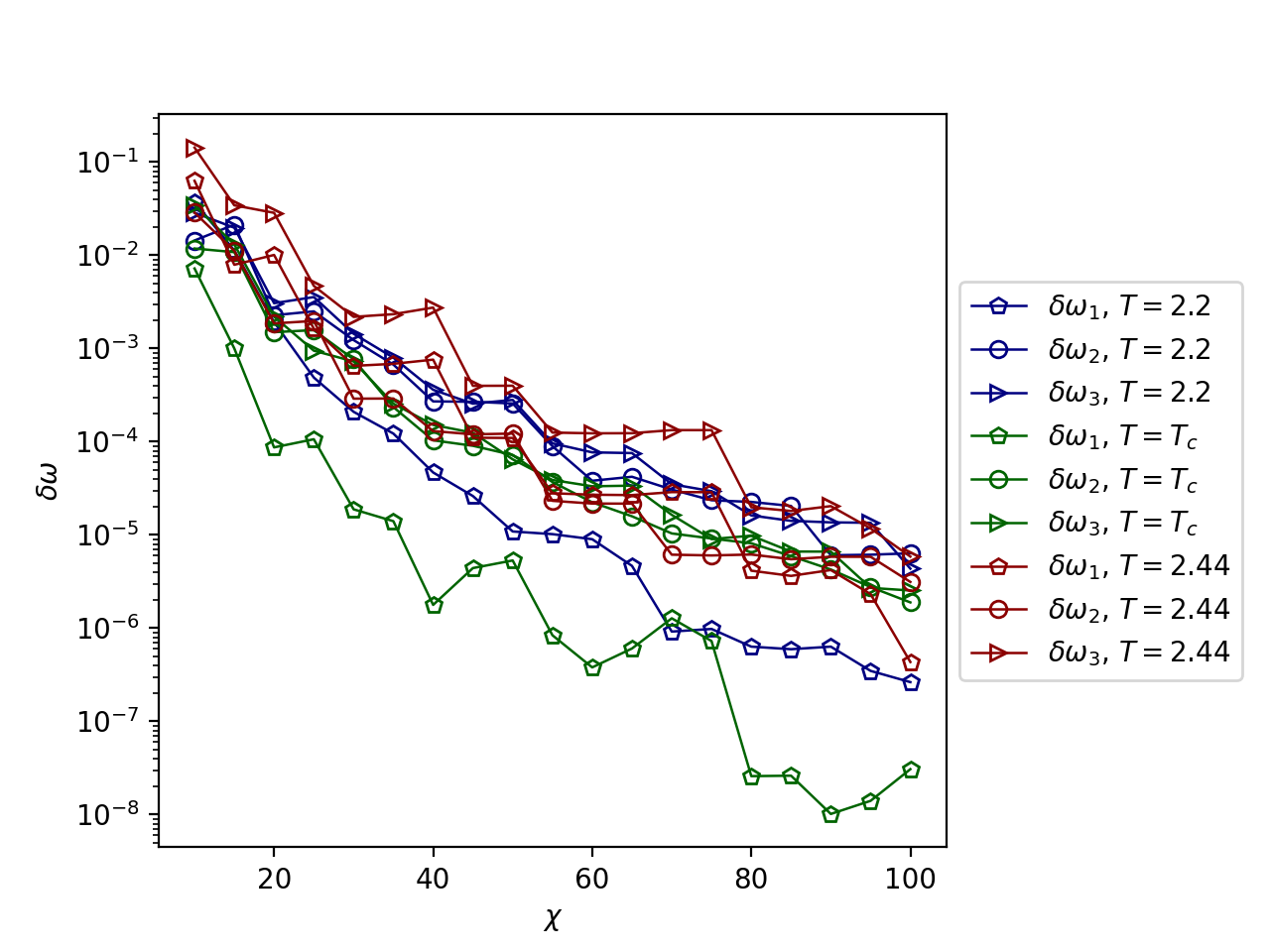}
\caption{The relative error of the three lowest energy gaps at three different temperatures $T=2.2,T_{\rm c}, 2.44$ with $L_x=2^6$ 
as a function of the bond dimension $\chi$.
}
\label{error_dcut}
\end{figure}

According to Sec.~\ref{sec:how_to}, we compute the energy gaps $\omega_a^{[{\rm hotrg}]}$ in Eq. (\ref{egaptn})
using HOTRG with a given bond dimension $\chi$.
Figure \ref{energy_spectrum}(a)
shows the three lowest energy gaps in the temperature range $T=2.2{-}2.44$ encompassing the critical point $T_{\rm c}=\frac{2}{\log(1+\sqrt{2})}$
for system size $L_x=2^5-2^7$ with $\chi=80$. 
We observe an expected behavior; by increasing the system size the lowest gap $\omega_1$ below $T_{\rm c}$ tends to be close to zero
while it stays nonzero for the temperature above $T_{\rm c}$.
In order to see an accuracy of the energy gap, we show its relative error,
\be
\delta\omega_a
=
\left|
\frac{\omega_a^{[\text{exact}]}-\omega_a^{[\text{hotrg}]}}{\omega_a^{[\text{exact}]}}
\right|,
\label{eqn:deltaomega}
\ee
in Fig.~\ref{energy_spectrum}(b). 
Here $\omega^{[\text{exact}]}_a$ is the Kaufman's exact results for finite volume \cite{PhysRev.76.1232}
(see Appendix \ref{sec:exact_spectrum} for details).
From this figure, we can see that the relative error increases for larger system size due to the iteration of the coarse-graining step.
We also observe that $\delta\omega_1$ 
apparently has a minimum around the critical point
while $\delta\omega_2$ and $\delta\omega_3$ do not show such a behavior. 
We note that the behavior of $\delta\omega_1$ is in contrast with the relative error of the free energy at finite volume,
\be
\delta f
=
\left|
\frac{f^{\text{[exact]}}_V-f^{[\text{hotrg}]}_V}{f^{[\text{exact}]}_V}
\right|,
\ee
where $f^{[\text{exact}]}_V$ is the exact free energy with volume $V=L^2$ with $L=L_x=L_\tau$ \cite{PhysRev.76.1232}. 
The relative error $\delta f$ is shown in Fig.~\ref{free_energy} where
the error becomes large around the critical point.

Next, let us see how the relative error of the energy gap scales with the bond dimension.
Figure \ref{error_dcut} shows $\delta\omega_a$ $(a=1,2,3)$ as a function of $\chi$
for selected values of the temperature.
From this figure, we can see that for all cases, the relative error tends to decrease when increasing the bond dimension.

\subsection{Quantum number classification}\label{sec:quantum_number}
 
\begin{figure}[t!]
\centering
\includegraphics[width=10cm,height=7.5cm]{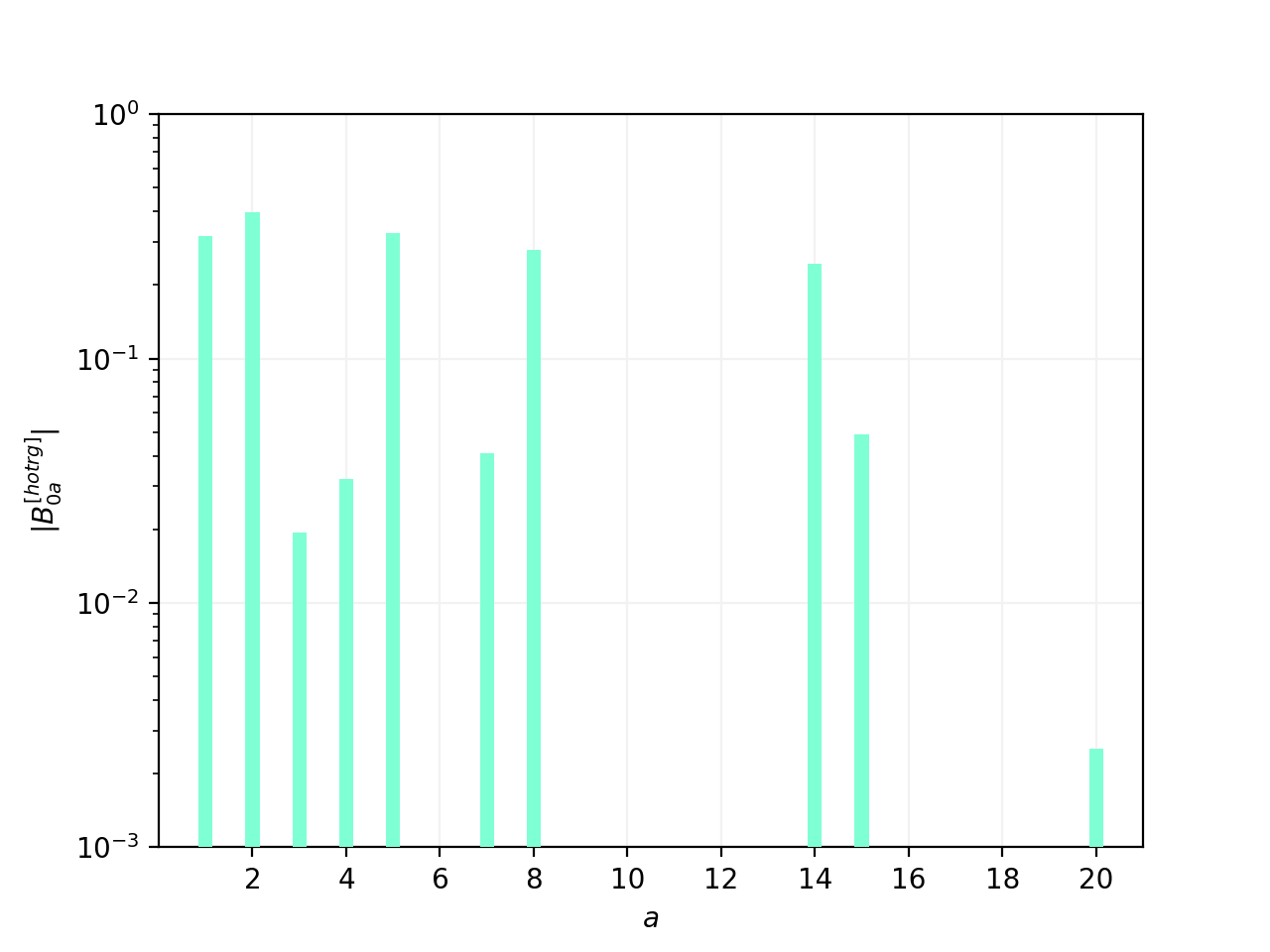}
\caption{The matrix element $B_{0a}^{[\text{hotrg}]}$ for $a=1,2,3,\hdots,20$
at $T=2.44$ with $L_x=2^6$ and $\chi=80$. 
}
\label{fig:B0a}
\end{figure}

\begin{table}[t!]
\begin{center}
\caption{Comparison to the exact result for $T=2.44$ and $L_{x}=2^{6}$.
$\delta\omega_a$ is the relative error of the energy gap 
defined in Eq. (\ref{eqn:deltaomega}).
The momentum $p$ is diagnosed in Sec.~\ref{sec:momentum}.
}
\label{t224d80}
\begin{tabular}{l|ll|ll|l|l}
\hline\hline
$a$ & $\omega_a^{[{\rm exact}]}$ &$q_a$&$\omega_a^{[{\rm hotrg}]}$  & $q_a$ &$\delta\omega_a$& $|p|$ \\ 
\hline
1	&0.1262302	&$-$	&0.1262307	&$-$	&0.000004 &$0$ \\
2	&0.1597880	&$-$	&0.1597889	&$-$	&0.000006 &$2\pi/L_x$\\
3	&0.1597880	&$-$	&0.1597911	&$-$	&0.000020 &$2\pi/L_x$\\
4	&0.2326853	&$-$	&0.2327046	&$-$	&0.000083 &$4\pi/L_x$\\
5	&0.2326853	&$-$	&0.2327095	&$-$	&0.000104 &$4\pi/L_x$\\
6	&0.2708016	&$+$	&0.2708359	&$+$	&0.000127 &$*$\\
7	&0.3181546	&$-$	&0.3183329	&$-$	&0.000560 &$6\pi/L_x$\\
8	&0.3181546	&$-$	&0.3183705	&$-$	&0.000679 &$6\pi/L_x$\\
9	&0.3290037	&$+$	&0.3291180	&$+$	&0.000347 &$*$\\
10	&0.3290037	&$+$	&0.3291425	&$+$	&0.000422 &$*$\\
11	&0.3290037	&$+$	&0.3291456	&$+$	&0.000431 &$*$\\
12	&0.3290037	&$+$	&0.3293794	&$+$	&0.001142 &$*$\\
13	&0.3872058	&$+$	&0.3878486	&$+$	&0.001660 &$*$\\
14	&0.4073042	&$-$	&0.4083937	&$-$	&0.002675 &$8\pi/L_x$\\
15	&0.4073042	&$-$	&0.4090231	&$-$	&0.004220 &$8\pi/L_x$\\
16	&0.4100181	&$+$	&0.4109090	&$+$	&0.002173 &$*$\\
17	&0.4100181	&$+$	&0.4112006	&$+$	&0.002884 &$*$\\
18	&0.4100181	&$+$	&0.4112120	&$+$	&0.002912 &$*$\\
19	&0.4100181	&$+$	&0.4114574	&$+$	&0.003510 &$*$\\
20	&0.4457831	&$-$	&0.4461242	&$-$	&0.000765 &$0$\\
\hline\hline
\end{tabular}
\end{center}
\end{table}

\begin{figure}[t!]
\centering
\includegraphics[width=10cm,height=7.5cm]{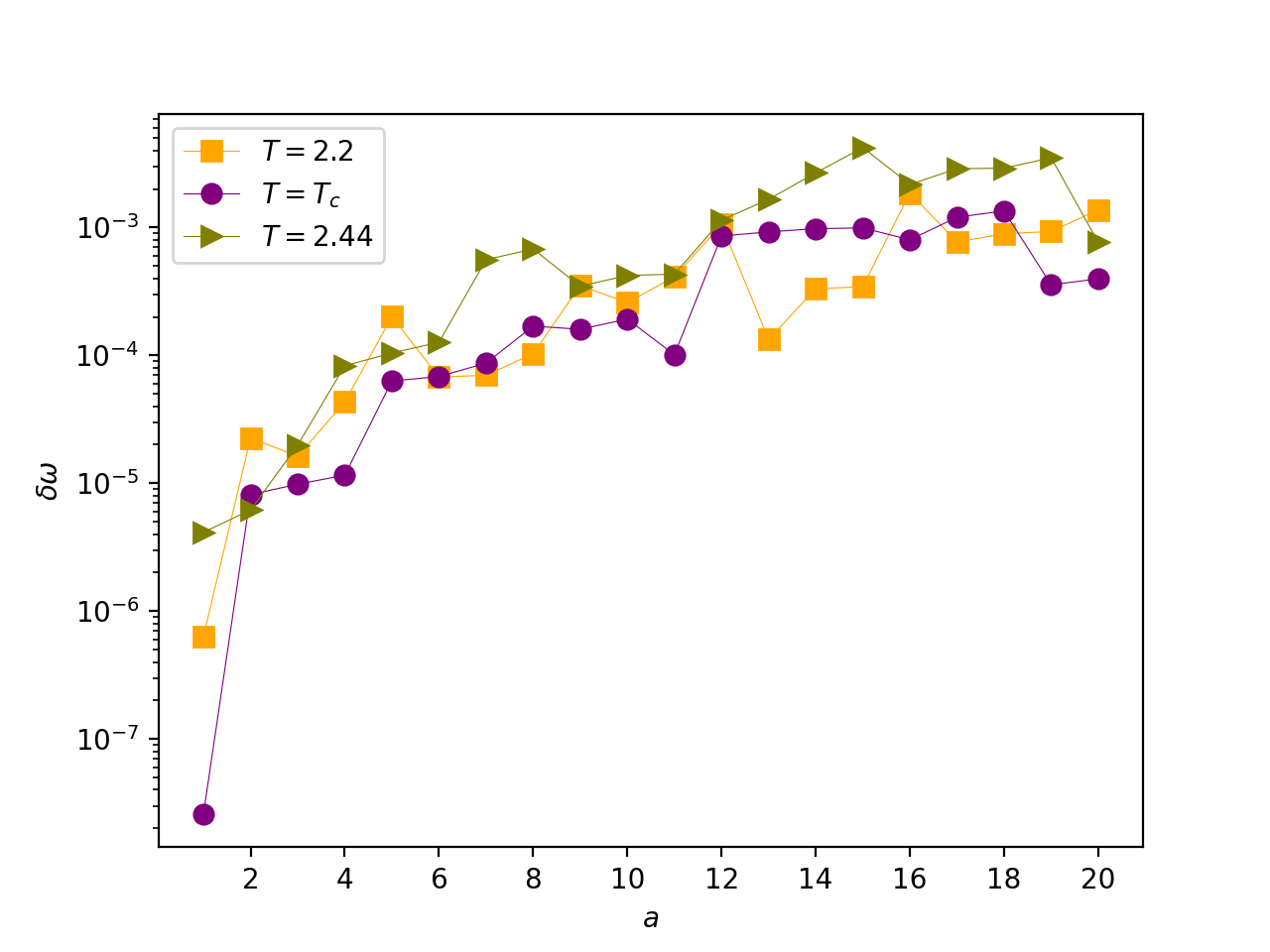}
\caption{The relative error of the energy gap $\delta\omega_a$ for $a=1,2,3,\hdots,20$ at three different temperature cases: $T=2.2$, $T_c$, and $2.44$. The system size is $L_x=2^6$ and the bond dimension is $\chi=80$.
}
\label{fig:error_enrg_a}
\end{figure} 

The Ising model with zero external magnetic field has $Z_2$ symmetry, thus
the energy eigenstates are divided into two groups labeled by the quantum number $q=\{+1,-1\}$.
In order to determine the quantum number of the eigenstates, 
following the procedure described in Sec.~\ref{sec:how_to}
we compute the matrix elements $B_{ba}^{[\text{hotrg}]}$ in Eq. (\ref{eqn:B_apx}) with a proper interpolation operator.
Here we choose the simplest choice i.e. a single spin field
$\mathcal{O}_q=s_x$ ($=s_0$ at $x=0$) whose quantum number is $q=-1$.

Once the elements of $B_{ba}^{[\text{hotrg}]}$ are estimated, the quantum number of the eigenstate can be determined from the selection rule as discussed in Sec.~\ref{sec:TM}.
Since the ground state ($a=0$) has quantum number $q=+1$,
we can classify the quantum number of the rest of the states labeled with $a=1,2,3,\hdots,$ by only looking at the first row of the estimated matrix $B_{0a}^{[\text{hotrg}]}\approx \langle \Omega |s_0|a\rangle$.
The selection rule tells us the quantum number of $|a\rangle$, $q_a$ as follows.
For each $a$,
\bea
\mbox{ if }
B_{0a}^{[\text{hotrg}]} \neq 0 &\Longrightarrow& q_a=-1.
\eea
To identify the states with $q_a=1$, we have to compute a matrix element using an operator whose quantum number is $q=1$.
However, the quantum number in the (1+1)d Ising model is restricted only either $-1$ or $1$.
Thus, knowing states with $q_a=-1$ will automatically determine the quantum number of the rest of the states, that is, $q_a=1$. 
See Fig.~\ref{fig:B0a} for the result of $B_{0a}^{[\text{hotrg}]}$ for $a=1,2,3,\hdots,20$ at $T=2.44$, $L_x=2^6$ and $\chi=80$.
The judged result of $q_a$ is listed in Table \ref{t224d80} together with the exact one obtained from Appendix \ref{sec:exact_spectrum}.
As a result, the quantum number is correctly judged up to 20 eigenstates for this parameter set.
On the other hand, due to the truncation error which is caused in the coarse-graining steps and strongly affects the high energy modes, our scheme fails to reproduce the correct quantum number for eigenstates with $a>20$ although we do not show them here.
In fact, it is difficult not only to judge the quantum number but also to obtain accurate energy gaps for higher excited states as seen in Table \ref{t224d80} (column for $\delta\omega_a$)
and Fig.~\ref{fig:error_enrg_a} where the relative error of the energy gap for $T=2.2$, $T_{\rm c}$ and $2.44$ tends to be large for larger $a$.

\subsection{Momentum identification}
\label{sec:momentum}

On a finite spatial volume,
the momentum is discretized as $p=2\pi k/L_x$ with $k=0,1,2,\hdots,L_x-1$
(or equivalently $k=-L_x/2+1,-L_x/2+2,\hdots,-1,0,1,\hdots,L_x/2-1,L_x/2$ assuming that $L_x$ is an even number),
and
the momentum of the single particle state can also be determined as follows.

\begin{figure}[t!]
\centering
\begin{subfigure}[b]{0.3\textwidth}
\centering
\includegraphics[width=5.5cm,height=5.5cm]{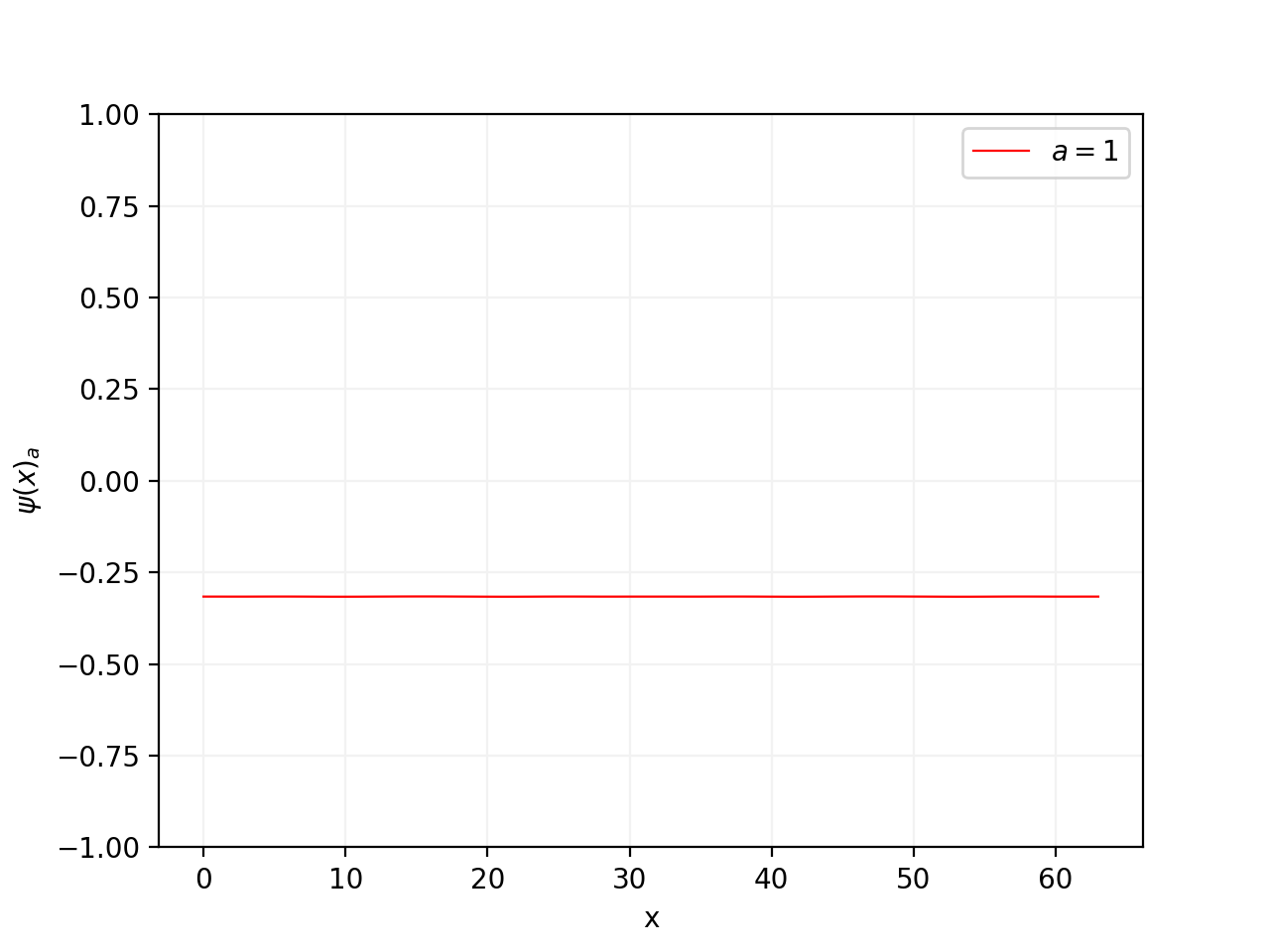}
\caption{}
\label{sfig:wf1}
\end{subfigure}\hspace{5mm}
\begin{subfigure}[b]{0.3\textwidth}
\includegraphics[width=5.5cm,height=5.5cm]{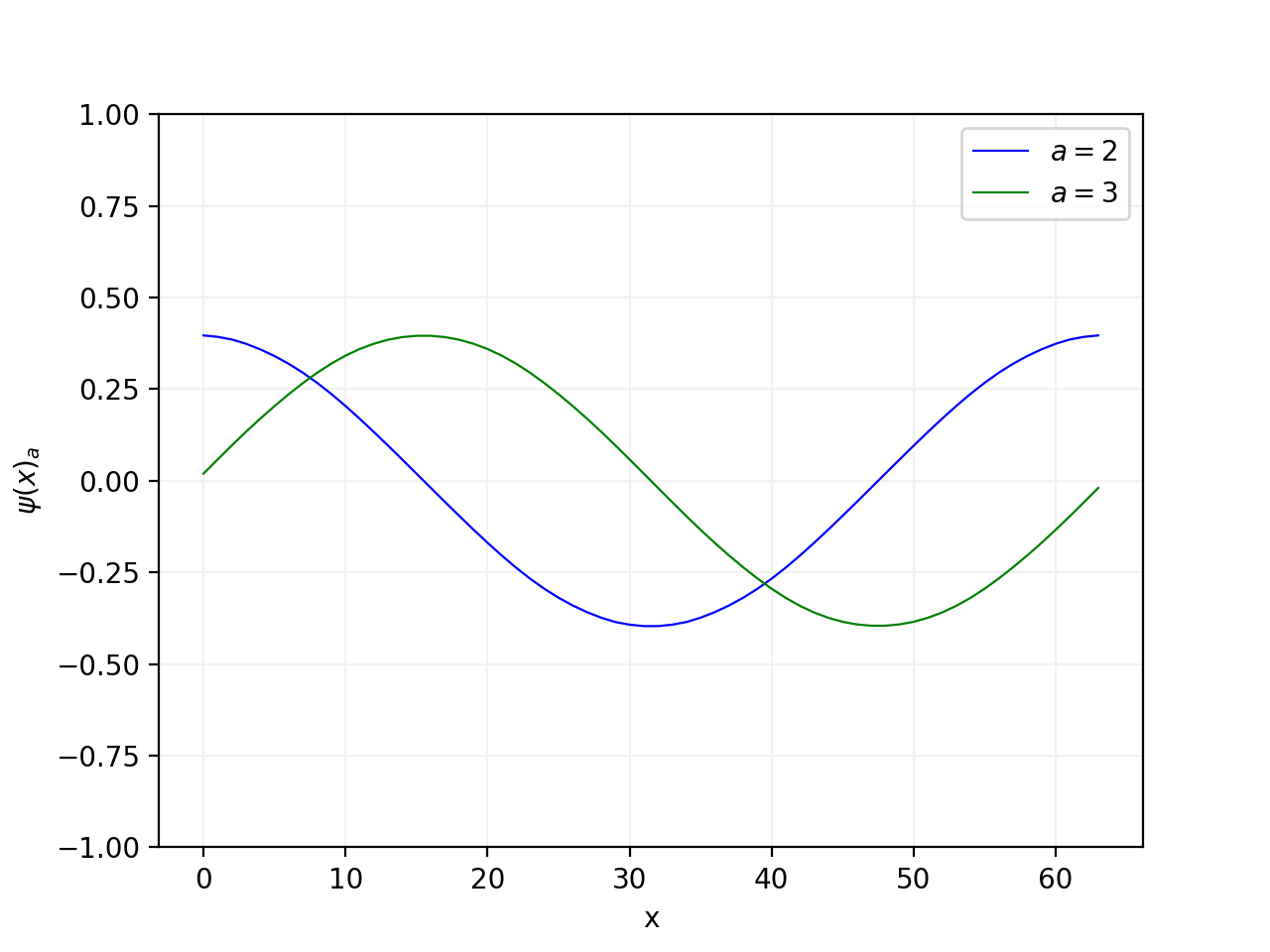}
\caption{}
\label{sfig:wf2}
\end{subfigure}\hspace{5mm}
\begin{subfigure}[b]{0.3\textwidth}
\includegraphics[width=5.5cm,height=5.5cm]{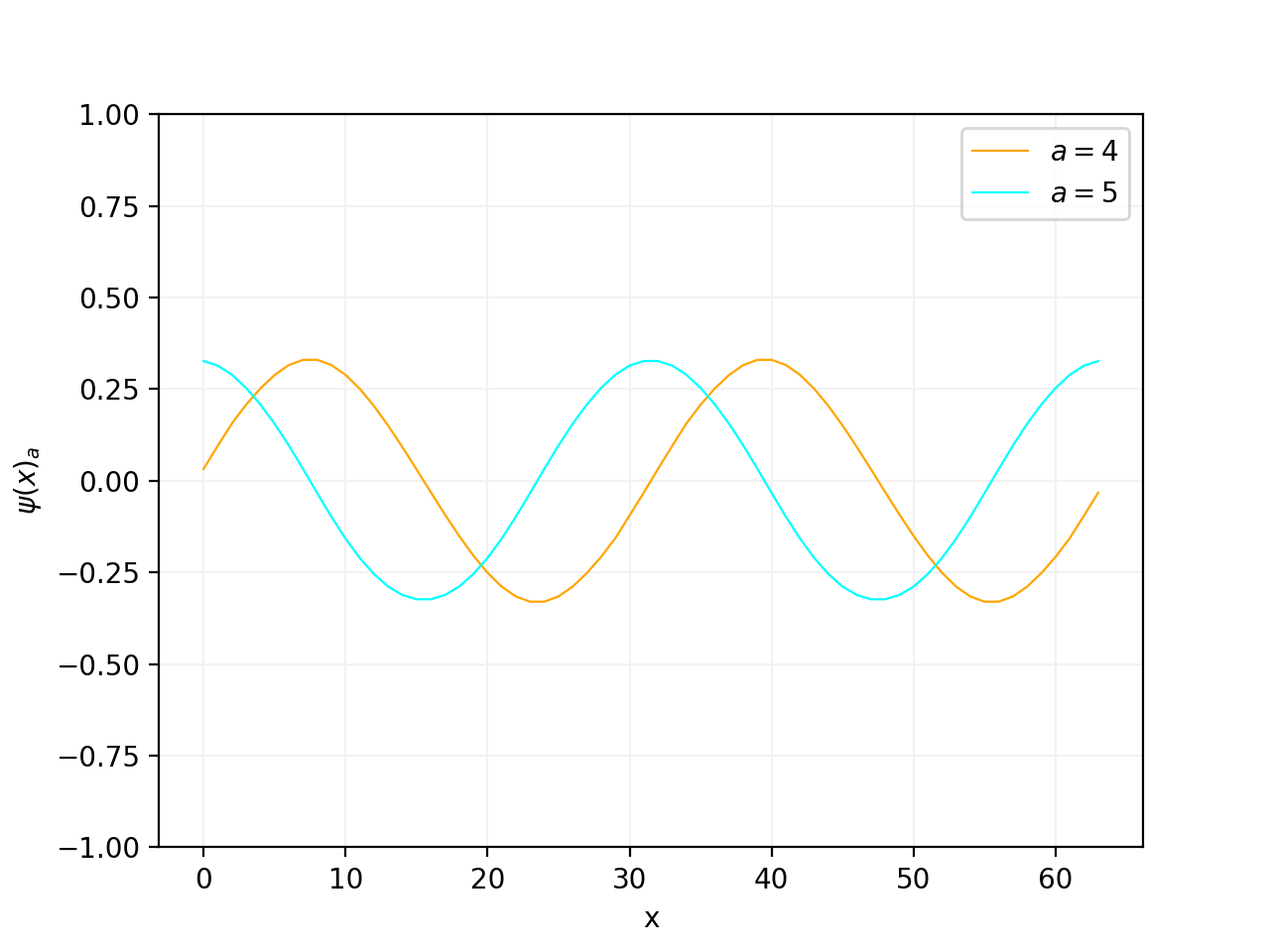}
\caption{}
\label{sfig:wf3}
\end{subfigure}\hspace{5mm}
\begin{subfigure}[b]{0.3\textwidth}
\includegraphics[width=5.5cm,height=5.5cm]{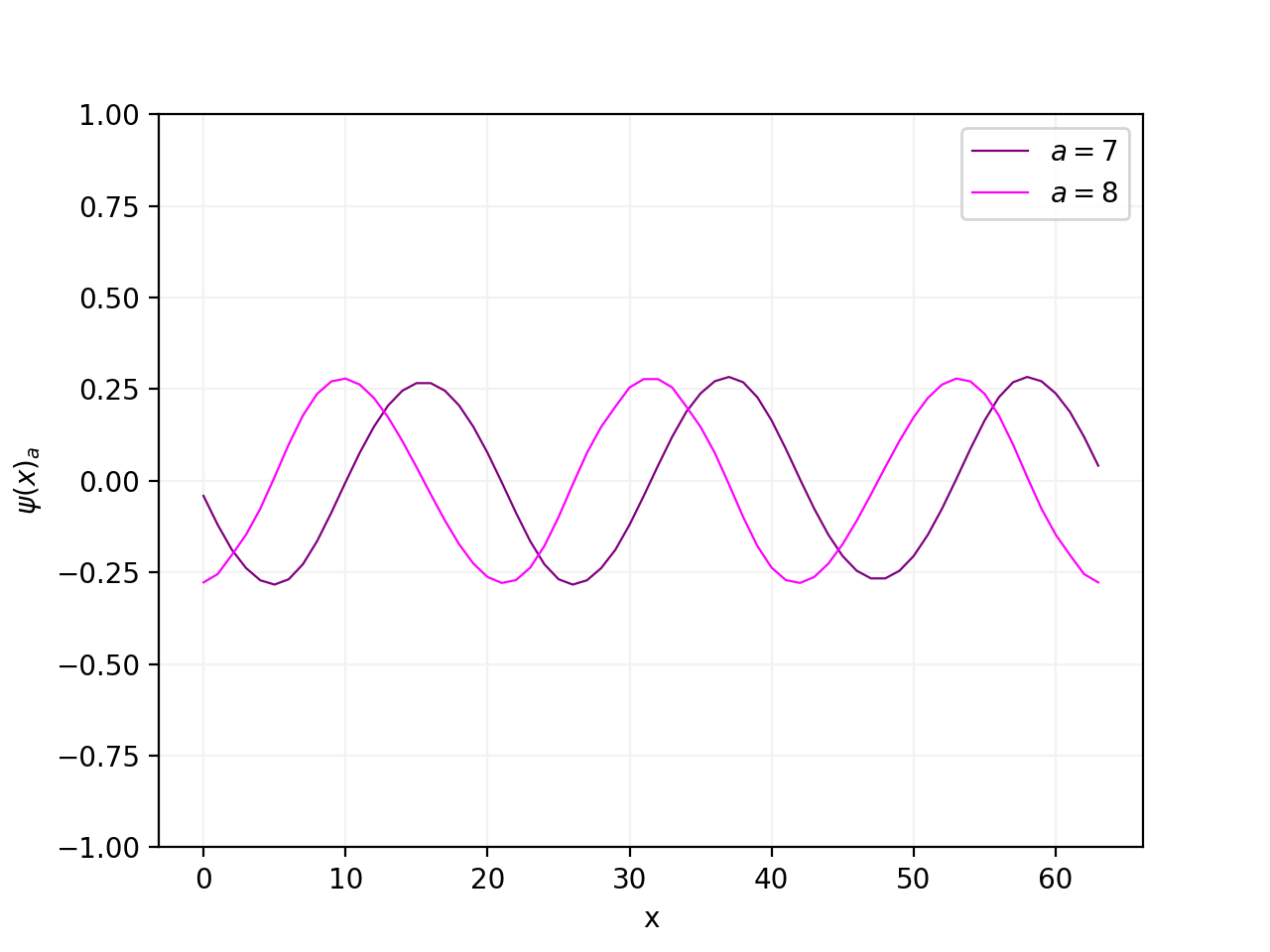}
\caption{}
\label{sfig:wf4}
\end{subfigure}\hspace{5mm}
\begin{subfigure}[b]{0.3\textwidth}
\includegraphics[width=5.5cm,height=5.5cm]{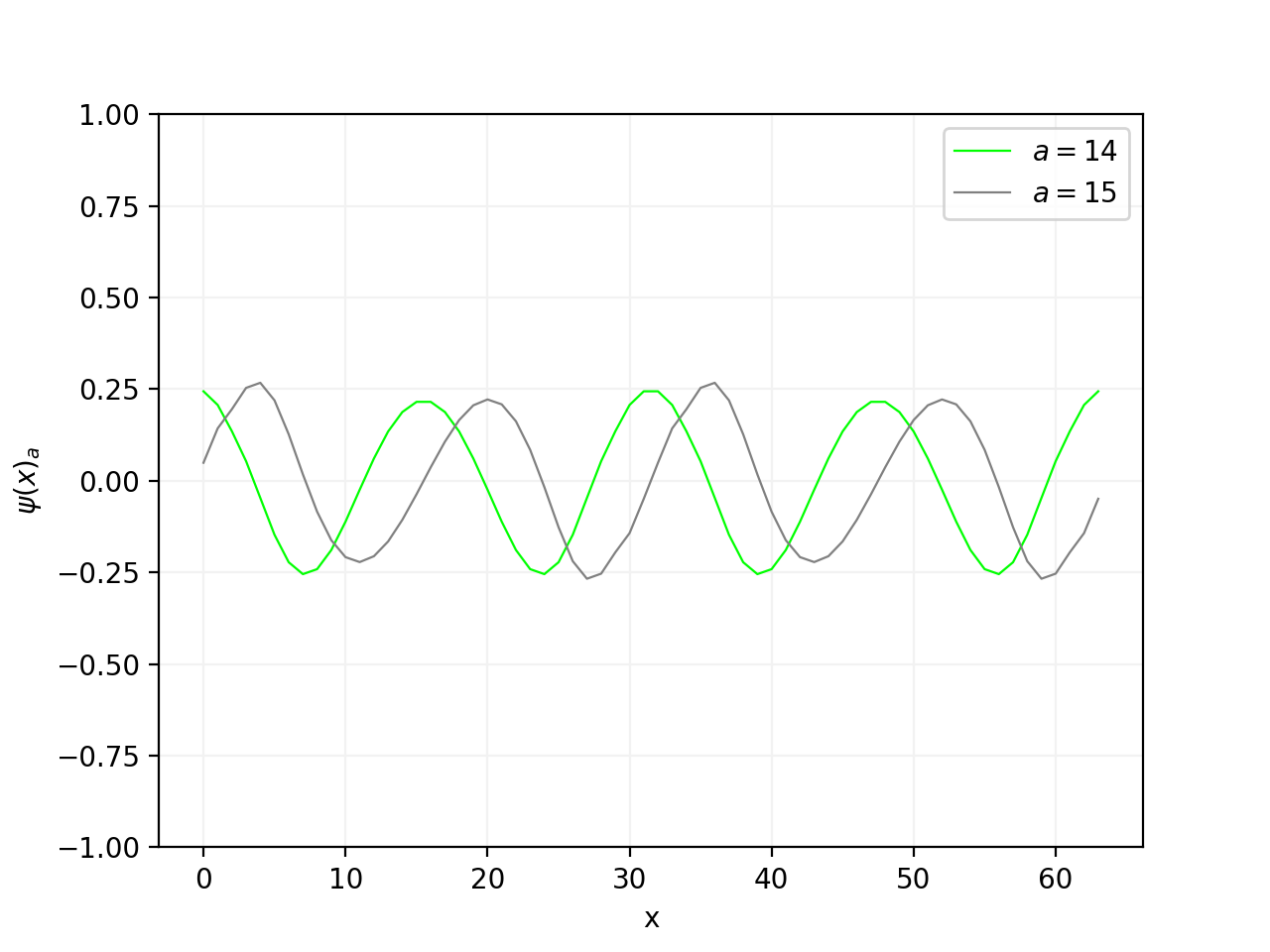}
\caption{}
\label{sfig:wf5}
\end{subfigure}\hspace{5mm}
\begin{subfigure}[b]{0.3\textwidth}
\includegraphics[width=5.5cm,height=5.5cm]{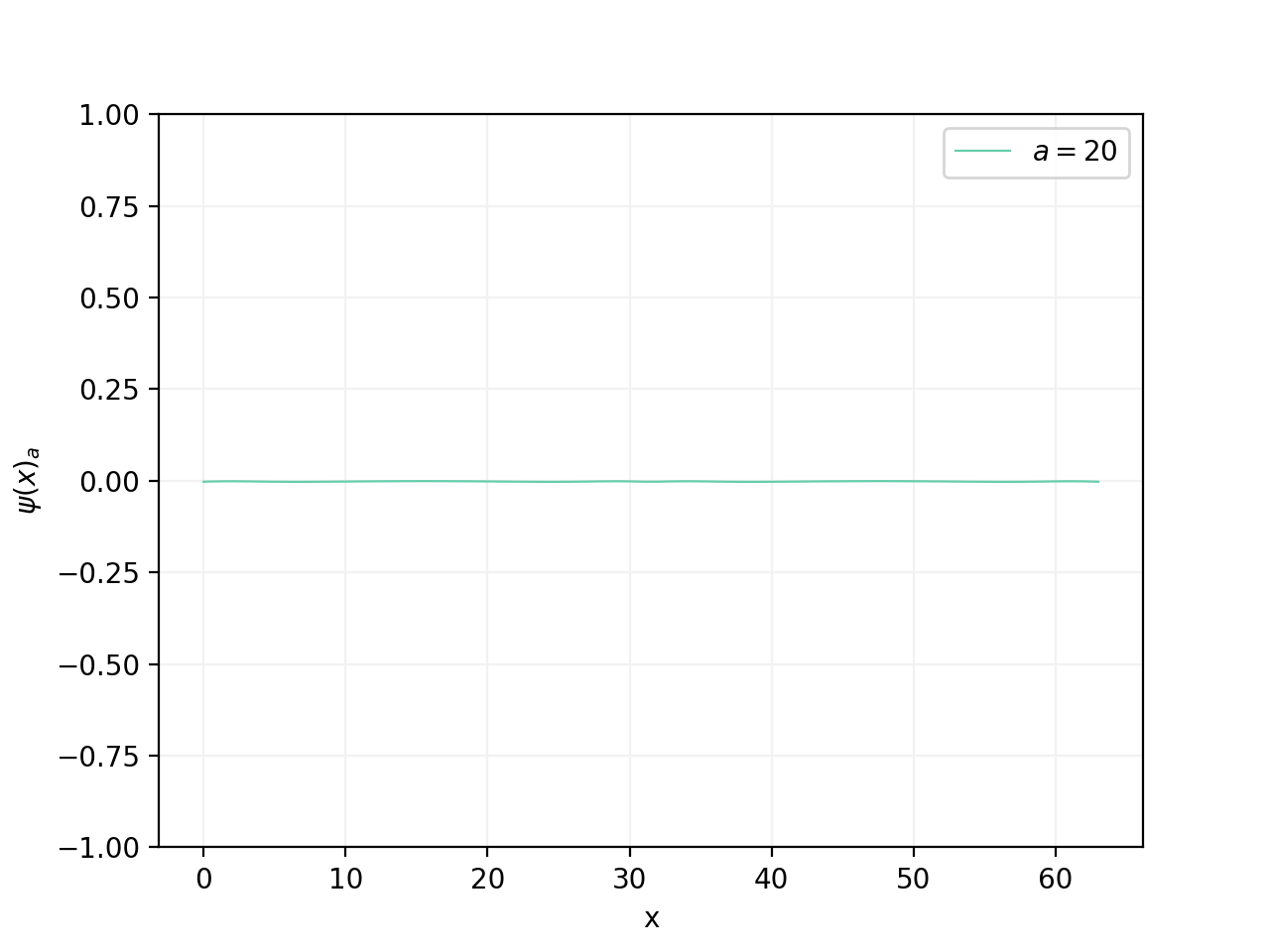}
\caption{}
\label{sfig:wf6}
\end{subfigure}
\caption{
The wave function of the eigenstates for the $q=-1$ sector
with $T=2.44$ and $L_x=2^6$ computed by HOTRG with $\chi=80$. 
The wave functions $\psi_a(x)$ for $a=1,2,3,4,5,7,8,14,15$, and $20$ are plotted in \subref{sfig:wf1}-\subref{sfig:wf6}.}
\label{fig:wavefunction}
\end{figure}

A simple way to check (the absolute value of) the momentum of a single particle state for the $q=-1$ sector is to look at its wave function in position space,
\be
\psi_a(x)=
\langle \Omega|s_x|a\rangle,
\ee
where $s_x$ is the spin field at $x=0,1,2,\hdots,L_x-1$.
The computation of the wave function can be done in a similar way to the matrix element
$B_{0a}^{[\text{hotrg}]}$ given in the previous subsection, but now we have to repeat it for
all possible values of $x$.
Figure \ref{fig:wavefunction} shows the numerical results of the wave functions for the $q=-1$ sector
($a=1,2,3,4,5,7,8,14,15,20$) with $T=2.44$, $L_x=2^6$, and $\chi=80$.
The wave function data is well described by functional form,
\be
\psi_a(x)\propto
\cos\left[p\left(x+\frac{1}{2}\right)\right]
\mbox{ or }
\sin\left[p\left(x+\frac{1}{2}\right)\right]
\ee
where $p=2\pi k/L_x$ is the discrete momentum.
For example, the $a=1$ state in Fig.~\ref{fig:wavefunction}(a) shows constant behavior; thus this is apparently a zero momentum state.
The states for $a=2$ and $3$ [see Fig.~\ref{fig:wavefunction}(b)] are described by
\be
\psi_2(x)\propto\cos\left[p\left(x+\frac{1}{2}\right)\right]
\mbox{ and }
\psi_3(x)\propto\sin\left[p\left(x+\frac{1}{2}\right)\right]
\mbox{ with }
p=\frac{2\pi}{L_x},
\ee
therefore the momentum of those states are judged to be 
$|p|=2\pi/L_x$.
The same thing can be applied to other states ($a=4,5$ states are paired and they correspond to momentum
$4\pi/L_x$,
and so on),
and the resulting
momentum is summarized in Table \ref{t224d80}.
One thing to be noted is that 
the wave function for $a=20$ in Fig.~\ref{fig:wavefunction}(f) seems to have zero amplitude over all $x$ at this scale of the $y$ axis,
but in fact it has nonzero amplitude of order $O(10^{-3})$ as seen in Fig.~\ref{fig:B0a}, where $\psi_{20}(0)$ is plotted,
therefore this is considered as a first excited state for the zero momentum channel.
The smallness of the amplitude simply reflects the small overlap of this state with the single spin field.


\begin{figure}[t!]
\begin{center}
\begin{tabular}{c}
\includegraphics[width=15cm,pagebox=cropbox,clip]{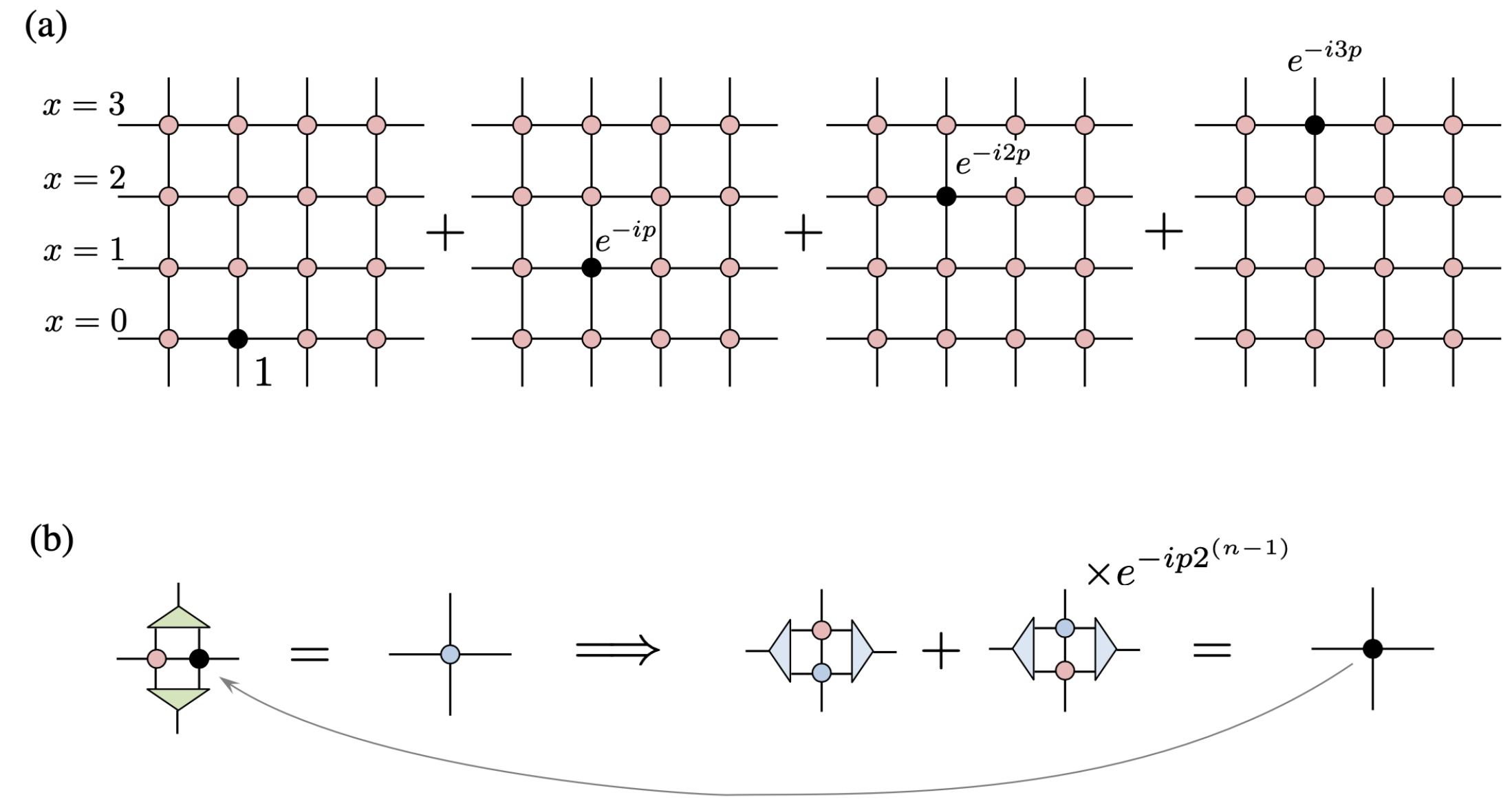}
\end{tabular}
\end{center}
\caption{
(a) Tensor network presentation with the impurity tensor
for the single field in the momentum space
$s_p=\frac{1}{L_x}\sum_{x}s_x e^{-ipx}$.
The black points represent the impurity tensor $A^\prime\times e^{-ipx}$
where $A^\prime$ is defined in Eq. (\ref{eqn:impurity_tensor}).
(b) A coarse-graining procedure for the tensor network of (a).
Here three-leg tensors represent the isometry for HOTRG and $n$ is the number of coarse-graining steps ($n=1,2,3,\hdots$).
}
\label{fig:coarse-graining_momentum}
\end{figure}

To quantitatively confirm the momentum identification presented in the previous paragraph,
we compute the matrix elements with a proper momentum field.
From the Fourier transformation of the spin field
\be
s_p
=
\frac{1}{L_x}
\sum_{x=0}^{L_x-1}
s_x
e^{-ipx}
\label{eqn:phip}
\ee
where the momentum is discretized $p=2\pi k/L_x$,
one can define matrix elements
\be
\langle\Omega|s_p|a\rangle
=
\frac{1}{L_x}
\sum_{x}
\langle \Omega|s_x|a\rangle e^{-ipx}.
\label{eqn:FT}
\ee
Using the matrix elements together with the selection rule, we can see that for given $a$ (in $q=-1$ sector) and $p$,
\be
\langle\Omega|s_p|a\rangle\neq 0
\Longrightarrow
|a\rangle\mbox{ belongs to } |p| \mbox{ sector}.
\ee
In this way, the momentum of $|a\rangle$ can be identified.
The matrix elements in Eq. (\ref{eqn:FT})
can be efficiently computed as shown in Fig.~\ref{fig:coarse-graining_momentum}
following the idea in \cite{MORITA201965}.
Table \ref{tab:momentum} shows numerical results of the matrix element
for $T=2.44$ and $L_x=2^6$.
In order to see the bond dimension dependence, we use $\chi=70,80,100$. 
For example, we can see that the momentum of $a=1$ and $20$ states is $p=0$,
and for $a=2$ and $3$, their momentum is $|p|=2\pi/L_x$, and so on.
We note that some states develop fake nonzero matrix elements due to the truncation error in the coarse-graining step.
We can, however, eliminate such a fake behavior by increasing the bond dimension.
For example, $a=20$ state has the nonzero matrix elements for $p=0,4\pi/L_x,8\pi/L_x$,
but the values for $p=4\pi/L_x$ and $8\pi/L_x$ tend to be small for larger $\chi$.
Thus, we conclude that the $a=20$ state belongs to the zero momentum.

\begin{table}[t!]
\begin{center}
\caption{The absolute value of the matrix elements $|\langle\Omega|s_p|a\rangle|$ 
for $T=2.44$ and $L_x=2^6$ using $\chi=70,80,100$.
Here we show only nonzero values that are larger than $O(10^{-5})$.
For the shaded case, 
the associated matrix elements are judged to be consistent with zero
since their values are small and they tend to be small for large $\chi$.
}
\label{tab:momentum}
\begin{tabular}{|l|l|r|l|l|l|}
\hline\hline
$k$&$|p|$&$a$&$\chi=70$&$\chi=80$&$\chi=100$\\
\hline
0&0&$1$&$0.31608$&$0.31609$&$0.31610$\\
& &$20$&$0.00195$&$0.00200$&$0.00208$\\
\hline
1&$2\pi/L_x$&$2$&$0.19850$&$0.19850$&$0.19850$\\
&&$3$         &$0.19844$&$0.19849$&$0.19850$\\
\hline
2&$4\pi/L_x$&$4$&$0.16406$&$0.16406$&$0.16409$\\
&&$5$         &$0.16370$&$0.16404$&$0.16406$\\
&&$\mathbf{1}$         &$\mathbf{0.00012}$&$\mathbf{0.00002}$&$\mathbf{<10^{-5}}$\\
&&$\mathbf{14}$        &$\mathbf{0.00001}$&$\mathbf{<10^{-5}}$&$\mathbf{<10^{-5}}$\\
&&$\mathbf{20}$        &$\mathbf{0.00021}$&$\mathbf{0.00017}$&$\mathbf{0.00012}$\\
\hline
3&$6\pi/L_x$&$7$&$0.13934$&$0.13940$&$0.13967$\\
&&$8$         &$0.13750$&$0.13932$&$0.13932$\\
&&$\mathbf{2}$         &$\mathbf{0.00006}$&$\mathbf{0.00006}$&$\mathbf{0.00002}$\\
&&$\mathbf{3}$         &$\mathbf{0.00061}$&$\mathbf{0.00010}$&$\mathbf{0.00006}$\\
\hline
4&$8\pi/L_x$&$14$  &$0.12013$&$0.12104$&$0.12201$\\
&&$15$&$<10^{-5}$&$0.12011$&$0.12012$\\
&&$\mathbf{1}$         &$\mathbf{0.00043}$&$\mathbf{0.00006}$&$\mathbf{0.00001}$\\
&&$\mathbf{4}$         &$\mathbf{0.00057}$&$\mathbf{0.00059}$&$\mathbf{0.00017}$\\
&&$\mathbf{5}$         &$\mathbf{0.00304}$&$\mathbf{0.00052}$&$\mathbf{0.00059}$\\
&&$\mathbf{19}$        &$\mathbf{0.11386}$&$\mathbf{<10^{-5}}$&$\mathbf{<10^{-5}}$\\
&&$\mathbf{20}$        &$\mathbf{0.00022}$&$\mathbf{0.00035}$&$\mathbf{0.00016}$\\
\hline
\end{tabular}
\end{center}
\end{table}

\begin{figure}[t!]
\centering
\includegraphics[width=10cm,height=7.5cm]{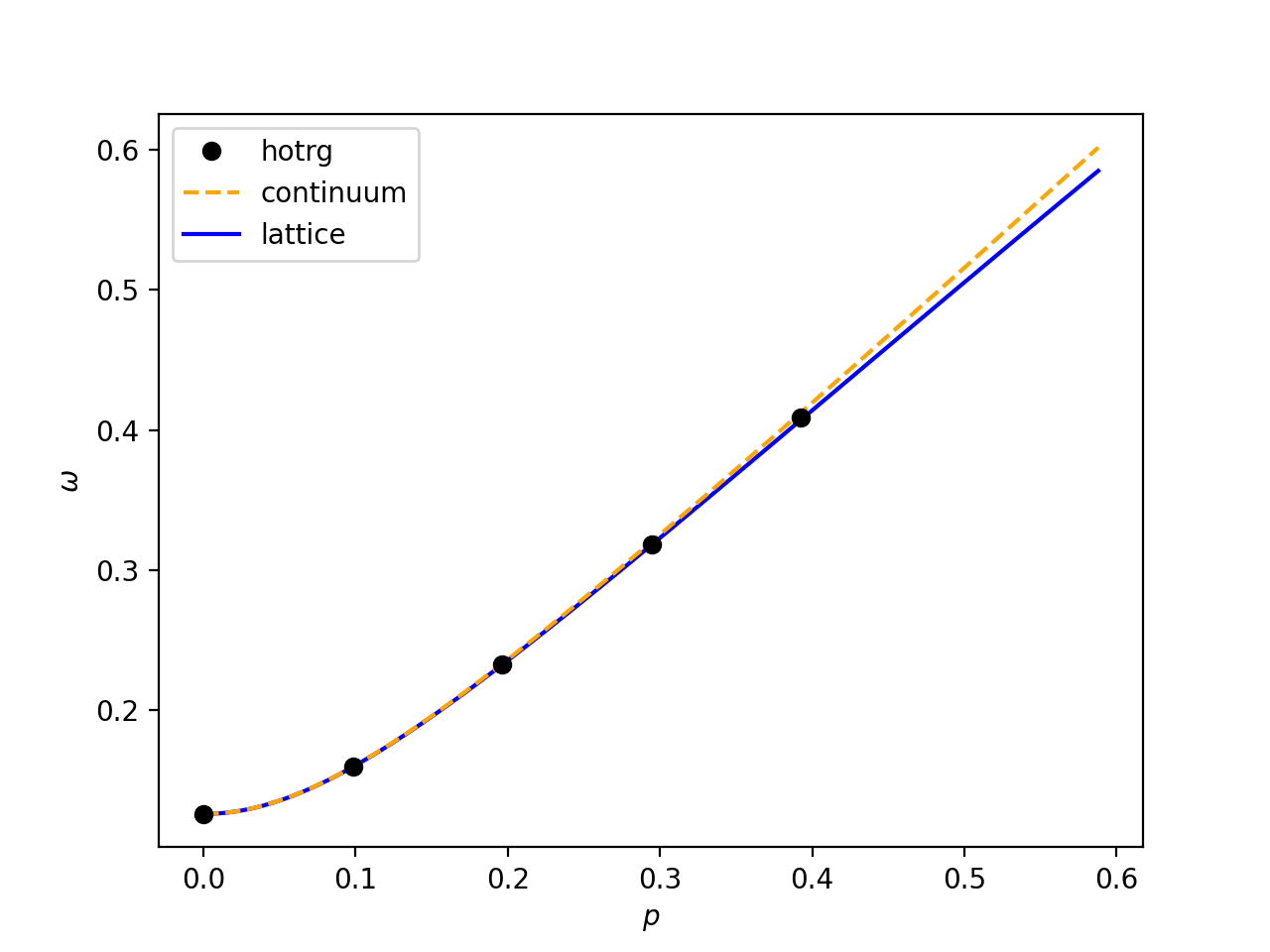}
\caption{The dispersion relation for the single particle of the $(1+1)$d Ising model at $T=2.44$ with $L_x=2^6$. The circle is the energy gap at the corresponding momentum computed by HOTRG with $\chi=80$. The dashed line is the continuum dispersion relation in Eq. (\ref{eq:disrel_continuum}) while the solid line is the lattice version in Eq. (\ref{eq:disrel_lat}).
}
\label{fig:disrel}
\end{figure}

As seen in the previous paragraph, the momentum is determined 
thus now we can check
the dispersion relation between the energy 
and the momentum. 
As seen in Table \ref{t224d80}, the degeneracy of the energy 
for the nonzero momentum with $q=-1$ is slightly broken
due to the truncation error, therefore
we use an average of them as the energy $\omega$.
See Fig.~\ref{fig:disrel} for the dispersion relation with $T=2.44$ and $L_x=2^6$.
The data points are generated by HOTRG with $\chi=80$ and compared with the continuum version of the dispersion relation
\be\label{eq:disrel_continuum}
\omega=\sqrt{m^2+p^2}
\ee
and the lattice version \cite{Gattringer:1992np}
\be\label{eq:disrel_lat}
\omega=\arccosh(1-\cos p + \cosh m)
\ee
where $m$ is the rest mass.
Both cases describe the data well and the lattice version is slightly better especially for the higher momentum region.

\subsection{Scattering phase shift}
\label{sec:phase_shift}

\begin{figure}[t!]
\begin{center}
\begin{tabular}{c}
\includegraphics[width=15.5cm,pagebox=cropbox,clip]{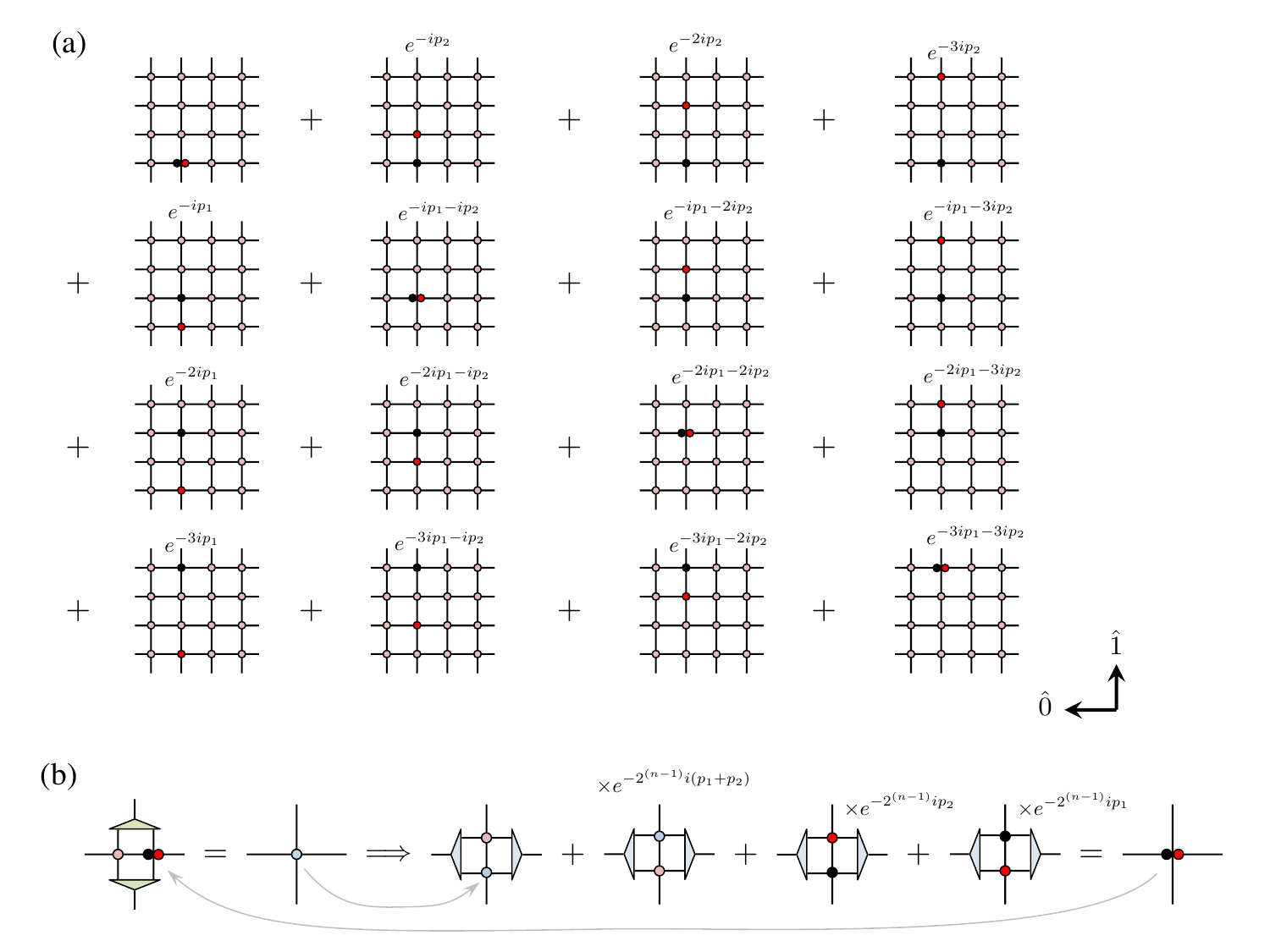}
\end{tabular}
\end{center}
\caption{
(a)Tensor network presentation with the impurity tensors
for the operator $\hat{\cal O}_2(P,p)$ in Eq. (\ref{eqn:two-particle_operator}).
The black (red) point represents the impurity tensor with the factor $e^{-ip_1x}$ ($e^{-ip_2x}$).
(b)A coarse-graining procedure for the tensor network of (a).
Here $n$ is the number of coarse-graining steps ($n=1,2,3,\hdots$).
}
\label{fig:coarse-graining_momentum_two_particle}
\end{figure}

\begin{figure}[h!]
\begin{center}
\begin{tabular}{c}
\includegraphics[width=12cm,pagebox=cropbox,clip]{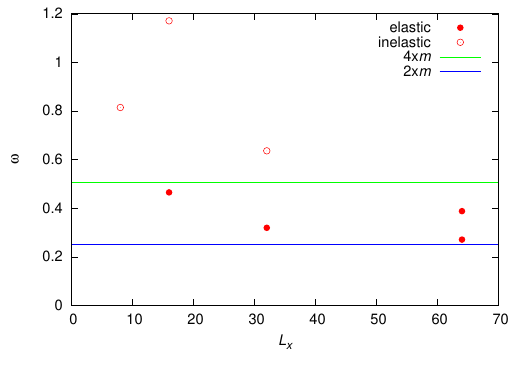}
\end{tabular}
\end{center}
\caption{
The two-particle state energy as a function of $L_x$.
Here $m$ is the rest mass for the one-particle state,
and we set the value in Eq. (\ref{eqn:rest_mass}).
}
\label{fig:w_L}
\end{figure}

In order to study the two-particle channel ($q=+1$ sector), we consider two-field operators
\be
{\cal O}_2(P,p)
=
\frac{1}{L_x^2}
\sum_{x,y=0}^{L_x-1}
s_xs_y
e^{-ip_1x-ip_2y}
\label{eqn:two-particle_operator}
\ee
where $p_1$ and $p_2$ are the discrete momentum $p_j=2\pi n_j/L_x$ with $n_j\in\mathbb{Z}$ ($j=1,2$),
and the total momentum $P$ and the relative $p$ are given by
\bea
P&=&p_1+p_2,
\\
p&=&\frac{1}{2}(p_1-p_2).
\eea
The matrix elements of the operators
\be
\langle\Omega|{\cal O}_2(P,p)|a\rangle
\label{eqn:matrix_element_two-field}
\ee
are useful to identify the momentum for the states with the $q=+1$ sector.
For example, for a given value of $P$, if the matrix element has a nonzero value,
\be
\langle\Omega|{\cal O}_2(P,p)|a\rangle
\neq0,
\ee
then the total momentum of $|a\rangle$ is estimated to be $P$ irrespective of $p$.
A tensor network representation with the impurity tensors for the two-field operator with $L_x=4$
is shown in Fig.~\ref{fig:coarse-graining_momentum_two_particle} where
its computational procedure is also described.
The numerical results of the matrix elements together with the corresponding energy for $T=2.44$, $L_x=2^3-2^6$, and $\chi=80$
are given in Table \ref{tab:total_zero_momentum_even_sector}.
The matrix elements of the zero total momentum operators with the states listed there have finite value and the same thing is confirmed even for large bond dimension $\chi=100$.
On the other hand, the matrix elements for finite total momentum are shown to be zero, therefore
we conclude that the states listed in the table belong to the zero total momentum sector.
See Fig.~\ref{fig:w_L} for the two-particle energy with the zero total momentum as a function of $L_x$.
\begin{table}[H]
\begin{center}
\caption{The tensor network results of the energy and the matrix elements in Eq. (\ref{eqn:matrix_element_two-field}) for some states with $q=+1$ sector at $T=2.44$, $L_{x}=2^3-2^{6}$, $\chi=80$.
The total momentum of all states listed here is judged to be zero.
}
\label{tab:total_zero_momentum_even_sector}
\begin{tabular}{rrrrrr}
\hline
$L_x$ & $a$ &$\omega_a^{[{\rm hotrg}]}$  & $\langle\Omega|{\cal O}_2(0,0)|a\rangle$ & $\langle\Omega|{\cal O}_2(0,2\pi/L_x)|a\rangle$& $\langle\Omega|{\cal O}_2(2\pi/L_x,\pi/L_x)|a\rangle$\\
\hline\hline
 8&  4& $0.814585$ & $0.37740$ & $0.12364$ & $<10^{-15}$ \\
  & 19& $2.133922$ & $0.07730$ & $0.04844$ & $<10^{-12}$ \\
\hline
16&  4& $0.465348$ & $0.31004$ & $0.09529$ & $<10^{-15}$ \\
  & 18& $1.171480$ & $0.06904$ & $0.05901$ & $<10^{-12}$ \\
\hline
32&  4& $0.319553$ & $0.21122$ & $0.06541$ & $<10^{-14}$ \\
  & 14& $0.636356$ & $0.04705$ & $0.06178$ & $<10^{-10}$ \\
\hline
64&  6& $0.270836$ & $0.12007$ & $0.03888$ & $<10^{-14}$ \\
  & 13& $0.387849$ & $0.03007$ & $0.05024$ & $<10^{-9}$ \\
\hline
\end{tabular}
\end{center}
\end{table}

\begin{figure}[t!]
\begin{center}
\begin{tabular}{c}
\includegraphics[width=12cm,pagebox=cropbox,clip]{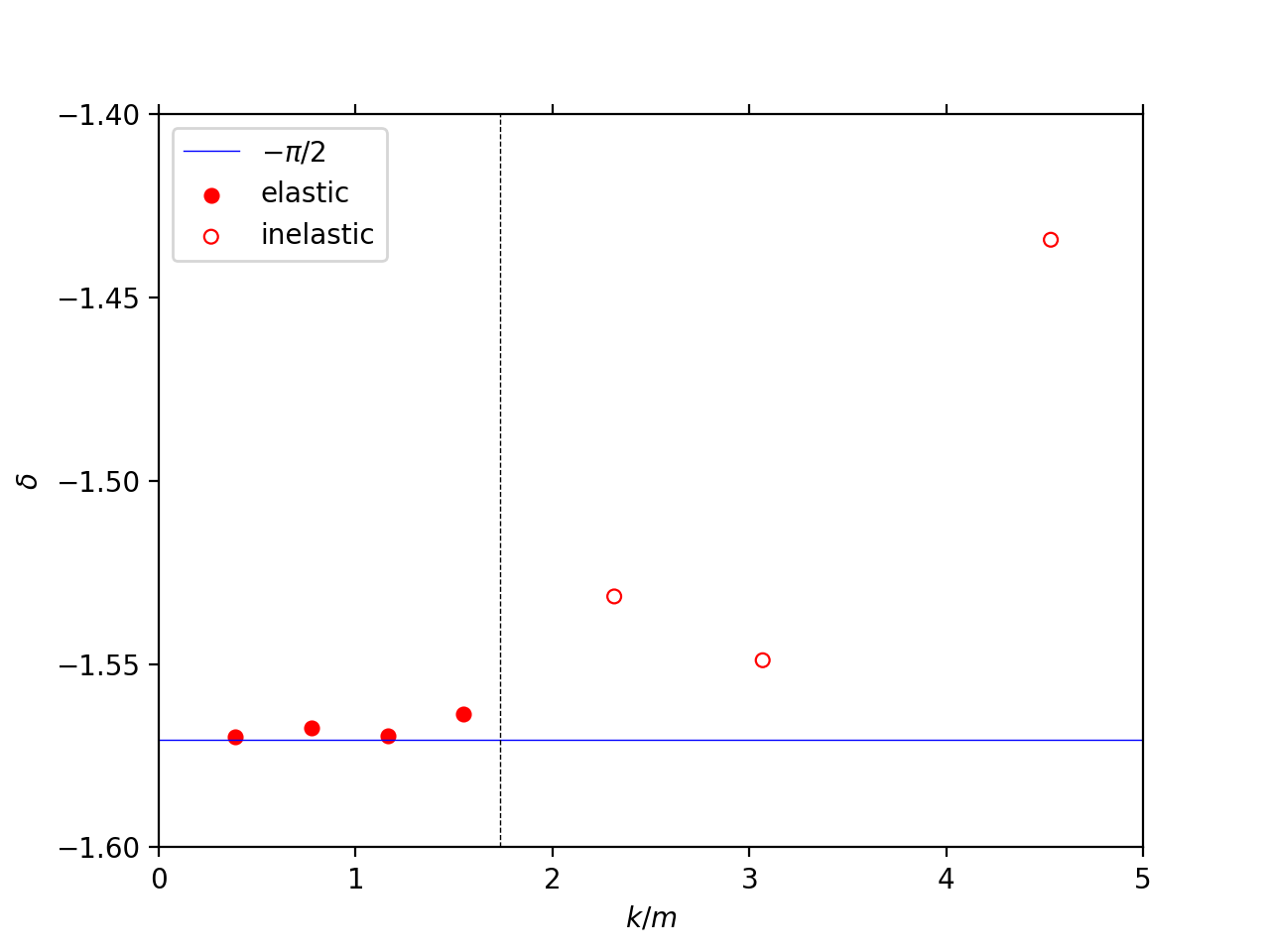}
\end{tabular}
\end{center}
\caption{
The phase shift as a function of $k/m$.
}
\label{fig:phase_shift}
\end{figure}

From the two-particle state energy $\omega$ in Table \ref{tab:total_zero_momentum_even_sector},
one can determine the relative momentum $k$, 
\be
\omega=2\sqrt{m^2+k^2},
\ee
where $m$ is the rest mass for one-particle state and here we set the exact value for infinite volume limit mass,
\be
m=0.12621870.
\label{eqn:rest_mass}
\ee
By using the value of $k$, the phase shift $\delta(k)$ is determined from L\"uscher's formula \cite{Luscher:1990ck},
\be
e^{2i\delta(k)}=e^{-ikL_x}.
\ee
See Fig.~\ref{fig:phase_shift} for the phase shift as a function of $k/m$.
The tensor network results in the elastic region which is defined as $2m\leq \omega < 4m$ or equivalently $0 \leq k/m<\sqrt{3}$ are consistent with the theoretical expectation $\delta(k)=\delta_{\rm Ising}=-\pi/2$ \cite{Gattringer:1992np}. 
On the other hand, the phase shift in the inelastic region is deviated from $-\pi/2$ since L\"uscher's formula is applicable only in the elastic region.

\section{SUMMARY}\label{sec:concl}

In the paper,
we have proposed a spectroscopy scheme by combining with the transfer matrix formalism and the Lagrangian tensor network formulation.
Using the new scheme, the energy spectrum can be simply obtained from the eigenvalues
of the numerical transfer matrix that is formed by the coarse-grained tensors.
The quantum number of the energy eigenstate is identified by the selection rule that
requires the matrix element of some insertion operator sandwiched by the eigenstates.
We have proposed the procedure to compute the matrix elements by using the impurity tensor network.

As a demonstration, we have studied the spectrum of the $(1+1)$d Ising model whose exact spectrum is well known.
We have confirmed that the energy spectrum and the quantum number are well reproduced up to 20 modes
for $L_x=2^6$ in disordered phase $T=2.44>T_{\rm c}$.
The accuracy of the lowest gap tends to be better around $T_{\rm c}$ in contrast to the free energy
where the accuracy gets worse around the critical point.
We have observed that the accuracy of the energy gaps tends to be worse for larger system size
due to the truncation error in the coarse-graining step, and
the systematic error becomes larger for higher excited states at the fixed system size.
On the other hand, for larger bond dimension, the accuracy of the energy gap gets smoothly better.
We also have computed the one-particle state wave function for the energy eigenstates and try to identify their momentum.
To confirm the momentum identification using the wave function, we have proposed another procedure using the selection rule
together with the proper matrix elements and then both results were shown to be consistent.
Relatively higher momentum states are properly identified and the dispersion relation is clearly observed.
We also identify the two-particle states and obtain the scattering phase shift from their energy using L\"uscher's formula.
The resulting phase shift is consistent with the theoretical expectation for the Ising model.

In the future, we plan to apply our new scheme to other quantum field theories.

\section*{ACKNOWLEDGEMENTS}
A. I. F. is supported by MEXT, JICA, and JST SPRING, Grant No. JPMJSP2135.
S.T. is supported in part by JSPS KAKENHI Grants No.~21K03531, and No.~22H01222.
T.Y. is supported in part by JSPS KAKENHI Grant No.~23H01195 and MEXT as ``Program for Promoting Researchers on the Supercomputer Fugaku'' (Grant No. JPMXP1020230409).
This work was supported by MEXT KAKENHI Grant-in-Aid for Transformative Research Areas A ``Extreme Universe'' No. 22H05251.

\appendix

\section{TRANSFER MATRIX AND INITIAL TENSOR FOR THE $(1+1)$d ISING MODEL}
\label{sec:TN_TM_2DIsing}

Consider the Ising model on the square lattice $\Gamma$ in Eq. (\ref{eqn:Gamma}) with 
the zero external magnetic field.
The Hamiltonian of the model is given by
\begin{equation}\label{hamilton}
H[s]=
-J\sum_{{\bm r}\in\Gamma}
\sum_{\mu=1}^2
s({\bm r}+\hat\mu)s({\bm r})
\end{equation}
where the spin variables take $s({\bm r})=\pm1$ and the interaction energy parameter $J$ is set to unity in the following.
The partition function for the model is given by
\begin{eqnarray}
Z=\sum_{\{s\}}e^{-\beta H}
\end{eqnarray}
with the inverse of temperature $\beta=T^{-1}$.
The periodic boundary condition is applied to the system. 

The partition function can be written in terms of transfer matrix 
\begin{eqnarray}
Z
&=&\Tr\left[\mathcal{T}^{L_{\tau}}\right]
\end{eqnarray}
where the transfer matrix ${\cal T}$ for the Ising model is given by
\bea
\mathcal{T}_{S^\prime S}&=&\left(\prod_{x=0}^{L_{x}-1}\exp
\left[\beta s(t+1,x)s(t,x)\right]\right)
\non\\
&\times&
\left(\prod_{x=0}^{L_{x}-1}
\exp\left[\frac{\beta}{2}s(t+1,x+1)s(t+1,x)
+\frac{\beta}{2}s(t,x+1)s(t,x)\right]\right).
\eea
The spin configurations on the Euclidean time slice at $t+1$ and $t$ are denoted by
\begin{alignat}{4}
&S^\prime&=&\{s(t+1&,&x)&|&x=0,1,2,\hdots,L_x-1\},
\\
&S&=&\{s(t&,&x)&|&x=0,1,2,\hdots,L_x-1\}.
\label{eqn:integrated_spin}
\end{alignat}

To derive an initial tensor for the Ising model, first we apply the EVD to the local Boltzmann factor,
\begin{equation}
e^{\beta s's}=\sum_{k=1}^2 u_{s^\prime k}\sigma_{k}(u^{\dagger})_{ks}
\,\,\,
\mbox{ for }
s^\prime,s=\pm1
\end{equation}
with
\begin{eqnarray}\label{matrx}
u_{sk}=\frac{1}{\sqrt{2}}\begin{pmatrix}
1&1\\
1&-1
\end{pmatrix},~~\sigma_{k}=\begin{pmatrix}
2\cosh \beta&0\\
0&2\sinh \beta
\end{pmatrix}.
\end{eqnarray}
By using the $u$ and $\sigma$, we define the initial tensor $A$ [see Eq. (\ref{eqn:initial_tensor}) for the scalar field case],
\be
A_{abcd}
=
\sqrt{\sigma_{a}\sigma_{b}\sigma_{c}\sigma_{d}}
\sum_{s=\pm1}
(u^\dag)_{as}
(u^\dag)_{bs}
u_{sc}
u_{sd}
\ee
where the indices $a,b,c,d$ take $1$ or $2$.
For a single spin field ${\mathcal{O}}_q=s$ with $q=-1$, the associated impurity tensor
[see Eq. (\ref{eqn:impurity_tensor}) for the scalar field case] is given by
\be
A_{abcd}^\prime
=
\sqrt{\sigma_{a}\sigma_{b}\sigma_{c}\sigma_{d}}
\sum_{s=\pm1}s
(u^\dag)_{as}
(u^\dag)_{bs}
u_{sc}
u_{sd}.
\ee

\section{EXACT SPECTRUM OF TRANSFER MATRIX FOR THE $(1+1)$d ISING MODEL}
\label{sec:exact_spectrum}
As given in \cite{PhysRev.76.1232}, 
for 
the inverse temperature $\beta$ and the spatial lattice size $L_x$,
the exact eigenvalues of the transfer matrix for the $(1+1)$d Ising model are given by
\begin{eqnarray}\label{lambda}
(2\sinh(2\beta))^{L_x/2}
\exp\left[
\frac{1}{2}
\left(
\pm\gamma_0\pm\gamma_2\pm\gamma_4\pm\hdots\pm\gamma_{2L_x-2}
\right)
\right]
\hspace{5mm}
&\mbox{for $-$ sector}&
\\
(2\sinh(2\beta))^{L_x/2}
\exp\left[
\frac{1}{2}
\left(
\pm\gamma_1\pm\gamma_3\pm\gamma_5\pm\hdots\pm\gamma_{2L_x-1}
\right)
\right]
\hspace{5mm}
&\mbox{for $+$ sector}&
\end{eqnarray}
where the even numbers of $+$ combination are only selected for each sector, thus there are in total $2\times 2^{L_x-1}$ eigenvalues.
Here $\gamma_n$ ($n=1,2,\hdots,2L_x$ and $\gamma_0=\gamma_{2L_x}$) is obtained by solving the equation
\bea
\cosh\gamma_n
&=&
\cosh(2\beta^\ast)\cosh(2\beta)
-
\cos\left(\frac{n\pi}{L_x}\right)
\sinh(2\beta^\ast)\sinh(2\beta),
\eea
where $\beta^\ast$ is defined from
$
e^{-2\beta}=\tanh\beta^\ast.
$

\section{SPECTROSCOPY OF ONE-DIMENSIONAL HOTRG}
\label{sec:1d_TN}

\begin{figure}[t!]
\begin{center}
\begin{tabular}{c}
\includegraphics[width=8cm,pagebox=cropbox,clip]{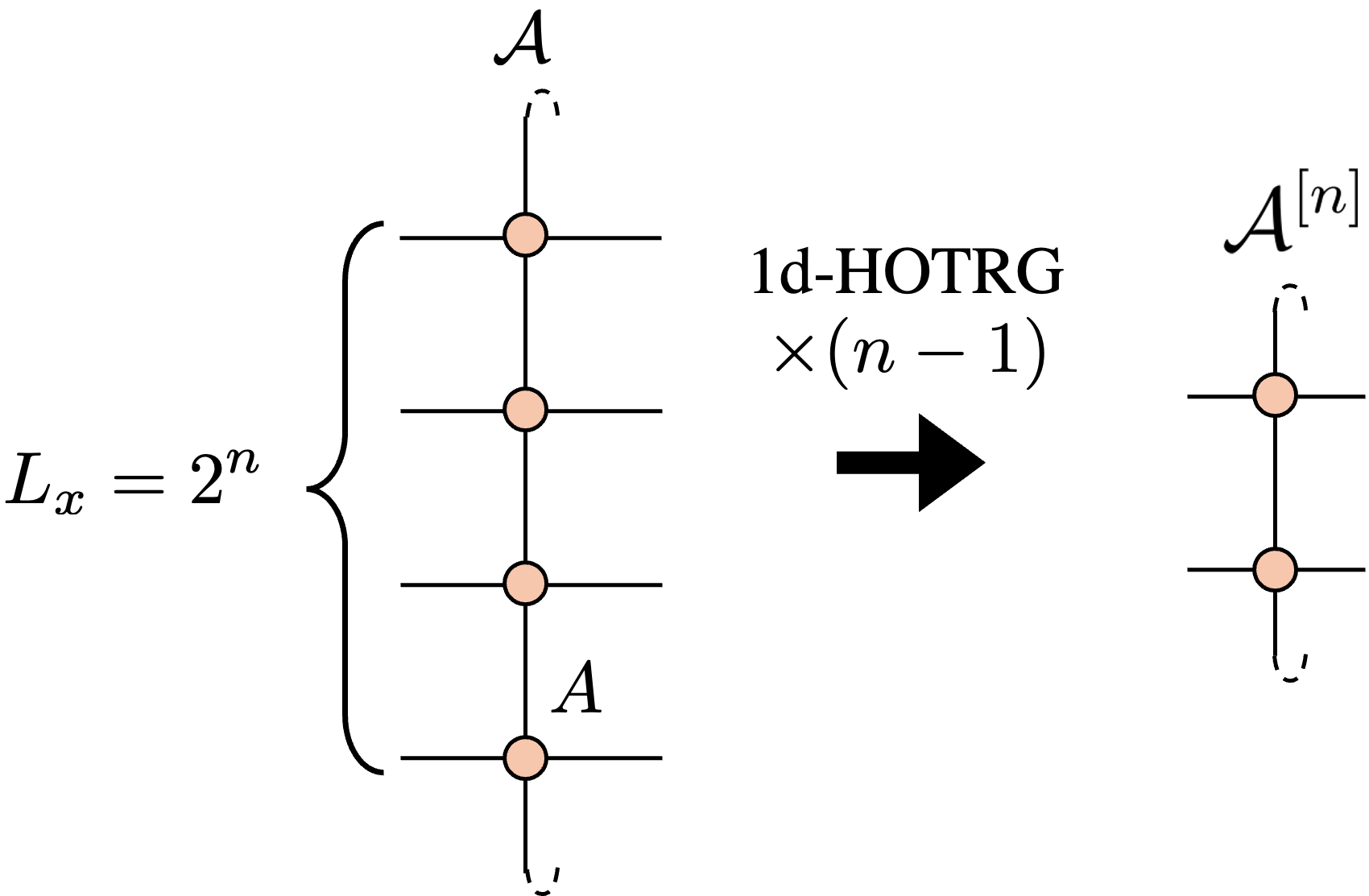}
\end{tabular}
\end{center}
\caption{
Coarse-graining process of one-time slice tensor network $\mathcal{A}$ by one-dimensional HOTRG.}
\label{fig:1dcoarse-graining}
\end{figure}

\begin{figure}[h!]
\begin{center}
\begin{tabular}{c}
\includegraphics[width=10cm,pagebox=cropbox,clip]{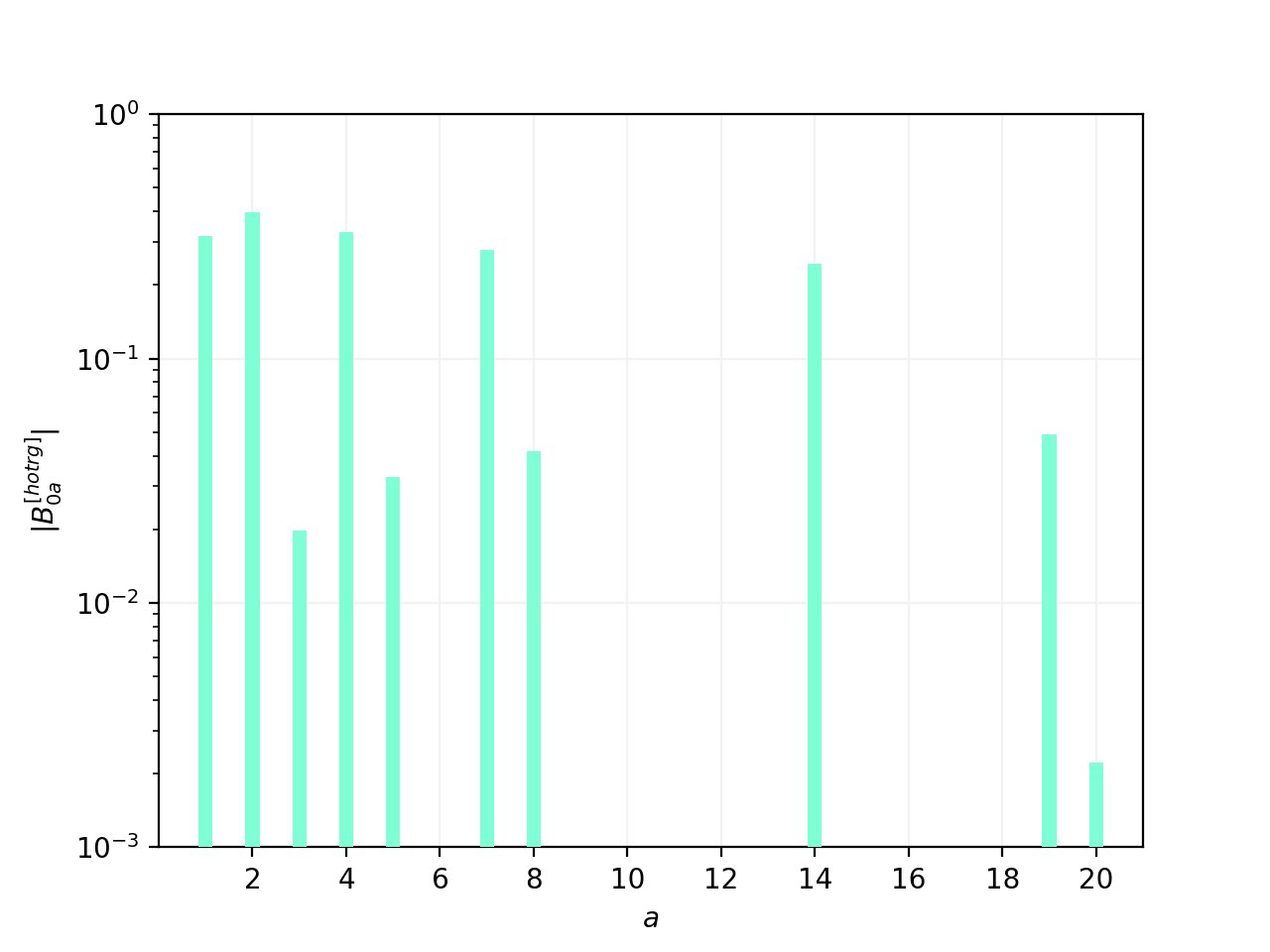}
\end{tabular}
\end{center}
\caption{
The matrix elements for the one-dimensional HOTRG at $T=2.44$ with $L_x=2^6$ and $\chi=80$.}
\label{fig:matrix_element_1d}
\end{figure}

In the body of the paper, we exclusively deal with the square tensor network for the spectroscopy.
To emphasize its effectiveness, here
we compare it with one-time slice tensor network $\mathcal{A}$ in Eq. (\ref{eqn:tildeT_A}) as shown in the left figure of Fig.~\ref{fig:1dcoarse-graining}.
By using the eigenvalues $\lambda$ and the unitary matrix $W$ of $\mathcal{A}$ in Eq. (\ref{eqn:tildeTM_EVD}),
the matrix element in Eq. (\ref{eqn:B}) can be expressed as
\be
B_{ba}
=
(U^\dag {\cal O}_q U)_{ba}
=
(
U^\dag {\cal T}^{-1} {\cal T}
{\cal O}_q {\cal T} {\cal T}^{-1} U
)_{ba}
=\lambda^{-1/2}W^{\dagger}\mathcal{A}'W\lambda^{-1/2}.
\label{eqn:B_one_time}
\ee
After coarse graining the network $\mathcal{A}$
using one-dimensional HOTRG as shown in Fig.~\ref{fig:1dcoarse-graining},
its diagonalization provides approximated $\lambda$ and $W$.
Approximated matrix elements are computed using Eq. (\ref{eqn:B_one_time}) with approximated
$\mathcal{A}^\prime$, $\lambda$ and $W$.
The relative error of the energy spectrum $\delta\omega_a$ for the Ising model is shown in Table \ref{1dvs2dhotrg}
where the standard two-dimensional HOTRG results are also shown for comparison.
As a result, the relative error of
the two-dimensional HOTRG is much smaller 
than that of the one-dimensional case.
The matrix elements are shown in Fig.~\ref{fig:matrix_element_1d} and the resulting quantum number judgment is listed in Table \ref{1dvs2dhotrg}.
Although the quantum number is correctly judged for the low lying states, some misjudgements occur for higher modes.

In conclusion, the spectroscopy using the one-time slice transfer matrix, which is obtained by one-dimensional HOTRG, is relatively affected by the coarse-graining truncation error than the two-dimensional case.
The reason is as follows.
The eigenvalue of the one-time slice transfer matrix tends to be degenerated especially for larger spatial size, therefore when coarse graining the transfer matrix,
the information compression does not work well and in the end such a large truncation error occurs.
On the other hand, for the square tensor network which is given by multiple products of the transfer matrix,
the degeneracy for the multiple transfer matrix is exponentially resolved and the information compression can work efficiently.


\begin{table}[H]
\begin{center}
\caption{The relative error $\delta\omega_a$ and the quantum number $q_a$ obtained from the one- and two-dimensional HOTRG
at $T=2.44$ with $L_x=2^6$ and $\chi=80$.}
\label{1dvs2dhotrg}
\begin{tabular}{|c|c|c|c|c|}
\hline\hline
$a$&$\delta\omega_a^{[\text{2d-hotrg}]}$&$q_a^{[\text{2d-hotrg}]}$&$\delta\omega_a^{[\text{1d-hotrg}]}$&$q_a^{[\text{1d-hotrg}]}$\\
\hline
1&0.000004&$+$& 0.002436&$+$\\
2&0.000006&$-$& 0.005464&$-$\\
3&0.000020&$-$& 0.007560&$-$\\
4&0.000083&$-$& 0.009567&$-$\\
5&0.000104&$-$& 0.012778&$-$\\
6&0.000127&$+$& 0.004938&$+$\\
7&0.000560&$-$& 0.013310&$-$\\
8&0.000679&$-$& 0.017084&$-$\\
9&0.000347&$+$& 0.029064&$+$\\
10&0.000422&$+$& 0.008621&$+$\\
11&0.000431&$+$& 0.008667&$+$\\
12&0.001142&$+$& 0.011308&$+$\\
13&0.001660&$+$& 0.012843&$+$\\
14&0.002675&$-$& 0.017259&$-$\\
15&0.004220&$-$& 0.020217&$+$\\
16&0.002173&$+$& 0.014054&$+$\\
17&0.002884&$+$& 0.014266&$+$\\
18&0.002912&$+$& 0.014409&$+$\\
19&0.003510&$+$& 0.015000&$-$\\
20&0.000765&$-$& 0.008715&$-$\\
\hline\hline
\end{tabular}
\end{center}
\end{table}

\bibliography{biblio}
\end{document}